\documentclass[conference,letterpaper,compsoc]{IEEEtran}
\usepackage[letterpaper, left=0.85in, right=0.85in, bottom=0.84in, top=0.61in]{geometry}

\usepackage{mathptmx} 

\newcommand{\ignore}[1]{}
\usepackage{fancyhdr}
\usepackage{cite}
\usepackage[normalem]{ulem}

\usepackage[bookmarks=true,breaklinks=true,letterpaper=true,colorlinks,linkcolor=black,citecolor=blue,urlcolor=black]{hyperref}

\usepackage[dvipsnames]{xcolor}
\usepackage{graphicx}
\usepackage{subfig}

\usepackage{mathptmx} 
\usepackage{fancyhdr}
\usepackage[normalem]{ulem}
\usepackage{colortbl}
\usepackage{enumitem}
\usepackage{lipsum}
\usepackage{amsmath}
\usepackage{tikz}
\usepackage{setspace}
\usepackage{cite}
\usepackage{soul}
\usepackage{tablefootnote}
\usepackage{threeparttable}

\usepackage{comment}
\usepackage{xspace}

\newcommand{\mechanismcap}[1]{DATAPLANT}
\newcommand{\mechanism}{Dataplant\xspace}
\newcommand{\mechanismtitle}[1]{Dataplant}

\newcommand{\uesa}[1]{US-Dataplant}
\newcommand{\upla}[1]{UC-Dataplant}
\newcommand{\dtran}[1]{D-Dataplant}

\newcommand{\uesaLong}[1]{\underline{U}npredictable Values in the DRAM \underline{S}As}
\newcommand{\uplaLong}[1]{\underline{U}npredictable Values in the DRAM \underline{C}ells}
\newcommand{\dtranLong}[1]{\underline{D}eterministic Values}

\newcommand{\uesaShort}[1]{US-Dplant}
\newcommand{\uplaShort}[1]{UC-Dplant}
\newcommand{\dtranShort}[1]{D-Dplant}
\newcommand{\UESAPUF}[1]{US-Dataplant PUF}
\newcommand{\UPLAPUF}[1]{UC-Dataplant PUF}
\newcommand*\circled[1]{\tikz[baseline=(char.base)]{
            \node[shape=circle,draw,inner sep=0pt,fill=black, text=white] (char) {#1};}}
            
\makeatletter
\newcommand\footnoteref[1]{\protected@xdef\@thefnmark{\ref{#1}}\@footnotemark}
\makeatother

\newif\ifsubmission
\submissionfalse

\ifsubmission
\newcommand{\lois}[1]{\textcolor{black}{#1}}
\newcommand{\minp}[1]{\textcolor{black}{#1}} 
\newcommand\nikaBox[1]{}
\newcommand\loiscomment[1]{}
\else
\definecolor{amber}{rgb}{1.0, 0.49, 0.0}
\newcommand{\lois}[1]{\textcolor{black}{#1}} 
\newcommand\loiscomment[1]{\noindent{\color{red}{\bf \fbox{Lois: }}{\it#1}}}
\newcommand{\minp}[1]{\textcolor{black}{#1}} 
\newcommand\nikaBox[1]{\noindent{\color{olive} {\fbox{\bf{Nika: }}\it"#1"}}} 

\fi

\expandafter\def\expandafter\UrlBreaks\expandafter{\UrlBreaks
  \do\a\do\b\do\c\do\d\do\e\do\f\do\g\do\h\do\i\do\j
  \do\k\do\l\do\m\do\n\do\o\do\p\do\q\do\r\do\s\do\t
  \do\u\do\v\do\w\do\x\do\y\do\z\do\A\do\B\do\C\do\D
  \do\E\do\F\do\G\do\H\do\I\do\J\do\K\do\L\do\M\do\N
  \do\O\do\P\do\Q\do\R\do\S\do\T\do\U\do\V\do\W\do\X
  \do\Y\do\Z}

\title{Dataplant: Low-cost In-DRAM Security Primitives \vspace{-1.2em}}
\title{\vspace{-3em}Dataplant: Enhancing System Security with\\ Low-Cost In-DRAM Value Generation Primitives\vspace{-1.5em}}

\newcommand{\affilETH}[0]{\textsuperscript{1}}
\newcommand{\affilNUDT}[0]{\textsuperscript{2}}
\newcommand{\affilSharif}[0]{\textsuperscript{3}}
\newcommand{\affilVU}[0]{\textsuperscript{4}}
\newcommand{\affilCMU}[0]{\textsuperscript{5}}
\newcommand{\affilMicrosoft}[0]{\textsuperscript{6}}
\newcommand{\affilHynix}[0]{\textsuperscript{7}}
\newcommand{\affilUnicamp}[0]{\textsuperscript{8}}

 \author{
 {Lois Orosa\affilETH}\quad%
 {Yaohua Wang\affilETH$^,$\affilNUDT}\quad%
 {Ivan Puddu\affilETH}\quad%
 {Mohammad Sadrosadati\affilETH$^,$\affilSharif}\quad%
 {Kaveh Razavi\affilETH$^,$\affilVU}\\%
 {Juan\ G\'omez-Luna\affilETH}\quad%
 {Hasan Hassan\affilETH}\quad%
 {Nika Mansouri-Ghiasi\affilETH}\quad%
 {Arash Tavakkol\affilETH}\quad%
 {Minesh Patel\affilETH}\\%
 {Jeremie Kim\affilETH$^,$\affilCMU}\quad%
 {Vivek Seshadri\affilMicrosoft}\quad%
 {Uksong Kang\affilHynix}\quad%
 {Saugata Ghose\affilCMU}\quad%
 {Rodolfo Azevedo\affilUnicamp}\quad%
 {Onur Mutlu\affilETH$^,$\affilCMU}\vspace{2pt}\\%
 {\small\it\affilETH ETH Z{\"u}rich  \qquad \affilNUDT National University of Defense Technology \qquad \affilSharif Sharif University of Technology  }
 \\
 {\small\it\affilVU Vrije  Universiteit Amsterdam \qquad \affilCMU Carnegie Mellon University \qquad \affilMicrosoft Microsoft \qquad \affilHynix SK Hynix \qquad \affilUnicamp UNICAMP}%
 \vspace{-1em}%
 }

\begin{document}
\setstretch{0.995}
\cfoot{\thepage}
\sloppy
\maketitle
\thispagestyle{plain}
\pagestyle{plain}

\begin{abstract}

\lois{
DRAM manufacturers have \minp{been} prioritizing memory capacity, yield, and bandwidth for years, while trying to keep the design complexity as simple as possible. 
DRAM chips are passive elements that store data, but they do not carry out any computation or other important function in the system, such as security.
%
Processors implement and execute most of the existing security mechanisms that protect the system against security threats, because 1) executing security mechanisms usually require non-trivial computational capabilities (e.g., encryption), and 2) commodity DRAM chips are not designed to perform computations or tasks other than data storage.}

\lois{In this work, we advocate for DRAM as a key component for providing security mechanisms to the system. 
%
 To this end, we propose \emph{\mechanism}, a new class of low-cost, high-performance, and reliable security primitives that can be integrated in commodity DRAM chips with minimal changes. The main idea of \mechanism{} is to slightly modify the internal DRAM timing signals to expose the inherent process variation found in all DRAM chips for generating \emph{unpredictable} but reproducible values (e.g., keys) within DRAM, without affecting regular DRAM operation.}

We use \mechanism{} to build two new security mechanisms. 
%
 \lois{First, a new \mechanism{}-based physical unclonable function (PUF)  with non-destructive read-out, low evaluation latency, robust responses,  resiliency to temperature changes, and   data-independent responses.}
\lois{Second,} a new cold boot attack prevention mechanism based on \mechanism{} that automatically destroys all data within DRAM on every power cycle with zero run-time energy and latency overheads. These mechanisms \lois{can be integrated} with current DDR memory \lois{chips without changing the DRAM array}. 

Using a combination of detailed simulations and experiments with \lois{136} real \lois{commodity} DRAM \lois{chips}, we show that our \mechanism{}-based PUF has \lois{1.8x} higher throughput than \lois{the best} state-of-the-art DRAM PUFs while being \lois{more} resilient to temperature changes, \lois{and totally independent of the values stored in memory}. We also demonstrate that our \mechanism{}-based cold boot attack protection mechanism is 19.5x faster and consumes 2.54x less energy when compared to existing mechanisms.


\end{abstract}

\vspace{-1mm}
\section{Introduction}


\lois{Modern processors have security support for encryption and memory isolation~\cite{costan2016intel} that protects secret data in memory from attackers\minp{. Unfortunately these mechanisms}  \minp{cause} significant performance and energy overheads~\cite{zhao2016performance}  and \lois{introduce new vulnerabilities}~\cite{brasser2017software, gotzfried2017cache}. Although DRAM is a key component \minp{of} many systems \minp{that often} stores critical or secret information, there is \minp{no} hardware security support implemented in commodity chips that accelerates or enhance\minp{s} security. }


\lois{We make a case for incorporating security primitives in commodity DRAM chips based on three fundamental observations. First, DRAM is ubiquitous in computer systems today, from high-end servers to low-cost Internet of Things (IoT) devices~\cite{HyperRAM}. Therefore, millions of users can benefit from having simple and low-cost security primitives in DRAM (e.g., easy to adopt by the industry). Second, critical data usually resides in main memory, and it is usually replicated across the processor caches. This introduces new sources for potential security breaches that make the system vulnerable to attacks (e.g., cache side channels). To reduce this security risks, one solution is to minimize \minp{data replication} across the system and process critical data close to where it resides (e.g., memory). Third, the data movement throughout the memory hierarchy causes  energy, performance and bandwidth overheads~\cite{mutlu2019processing}. }


\emph{Our goal} is to develop low-cost \lois{primitives in commodity DRAM chips} for supporting commonly-used security mechanisms. To this end, we propose \emph{\mechanism{}}, a novel low-cost, high-performance, and reliable \lois{class} of \lois{in-DRAM} security primitives. The key idea of \mechanism{} is to take advantage of inherent DRAM behavior to generate \emph{unpredictable values}, which we can use to support several \lois{system-level security mechanisms}. This work is the first to propose \lois{security primitives} that are simple enough so that can be \lois{integrated} with existing commodity DRAM chips. \lois{Our primitives are variants of existing DRAM commands (e.g., activation) that only require minimal timing changes \minp{to certain} internal DRAM signals. \minp{Therefore,} \mechanism{} primitives \minp{require \emph{no} changes to} the DRAM array. We propose two \mechanism{} primitives that complement each other: 1) \uesa{} generates values in the \lois{DRAM Sense Amplifiers (SAs), } and 2) \upla{} generates values in the DRAM cells.}

\minp{\uesa{} and \upla{}} \lois{enable} the implementation of security mechanisms efficiently in DRAM. We analyze and evaluate \minp{two such security mechanisms} in this work:
(1)~physical unclonable functions (PUFs)~\cite{Roel2012, Suh2007, Paral2011, Yu2012, Delvaux2015, Mandel2011, Lee2004, Selimis2011, Kang2014, Bhargava:2014, Roel2015, tuyls2006secret, yu2013security} and 
(2)~cold boot attack prevention~\cite{Yitbarek2017,Halderman2009,Simmons2011,Gruhn2013}.

\vspace{3pt}\noindent\textbf{1. Physical Unclonable Functions (PUFs).}
\lois{PUFs are usually used in cryptography to identify devices or to \lois{create} authentication keys. The main advantage of DRAM-based PUFs is that DRAM is ubiquitous in many computer systems today. However, existing DRAM PUFs~\cite{Tehranipoor:2015, Hashemian:2015, Keller2014, liu2014, Sutar2016, kim2018} have \lois{at least one of these} four main issues. First, \lois{most} DRAM PUFs have destructive read out (e.g., \lois{they destroy the memory content}). Second, \lois{most DRAM PUFs have a latency that} is very high to be used at runtime without system interference. Third, \lois{most} DRAM PUFs are very noisy, so they require a filtering mechanism to calculate the PUF responses.
Fourth, \lois{most of the DRAM PUFs provide responses that highly depend on temperature, which can affect the reliability of the generated keys. Avoiding this issue requires extra engineering effort (e.g., maintaining the DRAM at a constant temperature, or making the system temperature-aware). This issue is especially important in devices that are in the wild (e.g., IoT devices that \lois{use} DRAM~\cite{HyperRAM, apmemory, zentel})}.
In this work, we aim to solve these issues with our \mechanism{}-based DRAM PUFs (Section~\ref{sec:physicallyUnclonable}).
}



\vspace{3pt}\noindent\textbf{2. Cold Boot Attack Prevention.}
\lois{Physical attacks to computer systems are gaining relevance due to the fast growing of unsupervised systems in the wild (e.g., IoT devices) and the widespread use of  mobile systems (e.g., laptops).} 
One of the simplest and most effective physical attacks is \lois{the} cold boot attack~\cite{Halderman2009, Simmons2011, Gruhn2013, Hilgers2014, BAUER2016S65,Yitbarek2017}. The goal of a cold boot attack is to retrieve secret information stored in DRAM from the victim's computer system. A common approach to mitigate this attack is by encrypting the \emph{entire} memory~\cite{Suh:2003, arnold2004ibm, Yang:2005, Duc2006, Rogers2007, Henson:2014}, which can be expensive in terms of hardware \lois{cost}, power consumption and performance.
\lois{In this work, we propose a cold boot attack prevention mechanism based on \mechanism{} that} immediately and reliably destroys all data inside DRAM, automatically at power up, without incurring any latency or power overhead at runtime (Section~\ref{sec:coldbootattacks}).

We extensively evaluate \lois{our \mechanism{} primitives} and the two security \lois{mechanisms that they enable by} using a combination of detailed circuit-level simulations, system simulations, and experiments on \emph{real} DRAM chips. We obtain two key results. First, our Dataplant-based PUFs are \lois{1.8x faster} while achieving  better repeatability under changing temperatures \lois{than the best state-of-the-art DRAM PUFs~\cite{kim2018,talukder2019prelatpuf}}, \lois{and being totally independent of the values stored in memory}. We also show that the unpredictable numbers generated by \mechanism{} \lois{with data obtained from real commodity DRAM chips} pass \lois{all NIST tests ~\cite{rukhin2001statistical}}, demonstrating randomness and suitability for security mechanisms. Second, our Dataplant-based cold boot attack protection is 34x faster and 2.5x more energy efficient than a system implementing our prevention mechanism without \mechanism{}.  \mechanism{} also enables other security mechanisms (e.g., secure deallocation~\cite{chow2005shredding}), and it can be used by system designers and software developers once \mechanism{} is available in commodity DRAM chips. 


We make the following \lois{key} contributions:

\begin{itemize}[leftmargin=3mm,itemsep=0mm,parsep=0mm,topsep=0mm]
\item \lois{We propose \mechanism{}, a new class of low-cost in-DRAM primitives that  \lois{generate values in DRAM \lois{(Section~\ref{sec:dataplant})}. \mechanism{} enables} new security mechanism\minp{s} in all systems that use \minp{commodity} DRAM. \lois{We design two different \mechanism{} implementations with different trade-offs}. \mechanism{} is especially practical in computer systems with limited hardware or low-cost requirements.}


\item \lois{We propose two new DRAM PUFs based on \mechanism{} primitives that solve some of the main issues of state-of-the-art DRAM PUFs (Section~\ref{sec:physicallyUnclonable}).}

\item \lois{We propose a new cold boot attack prevention mechanism based on \mechanism{} primitives that does not incur any latency or power overhead at runtime (Section~\ref{sec:coldbootattacks}).}

\item We extensively evaluate \lois{our \mechanism{} \minp{primitives} and the security mechanisms they enable}, and \minp{we} demonstrate that \lois{they} are significantly faster and more energy-efficient than their state-of-the-art counterparts \lois{(Section~\ref{sec:evaluation})}.

\end{itemize}

\section{Background}
\label{sec:background}

\subsection{DRAM}
\label{sec:dramorganization}
We provide background on the   DRAM architecture relevant to this work. We describe the organization of a DRAM chip, the architecture of its sense amplifiers and the operations that are performed on a DRAM chip. 

\vspace{5pt}\noindent\textbf{DRAM Organization} DRAM chips are manufactured in a variety of configurations~\cite{jedec2012jedec}, including a range of capacities and data bus widths ranging between 4 and 16 pins. Since an individual DRAM chip has a small capacity and a limited data width, multiple DRAM chips are usually grouped together in the same DRAM module to form a \emph{rank}, providing a larger data bus (usually 64-bits wide). Specialized DRAM for IoT can have fewer chips and a narrower bus~\cite{ HyperRAM, apmemory, zentel}. 

\begin{figure}[h] \centering
    \includegraphics[width=1.0\linewidth]{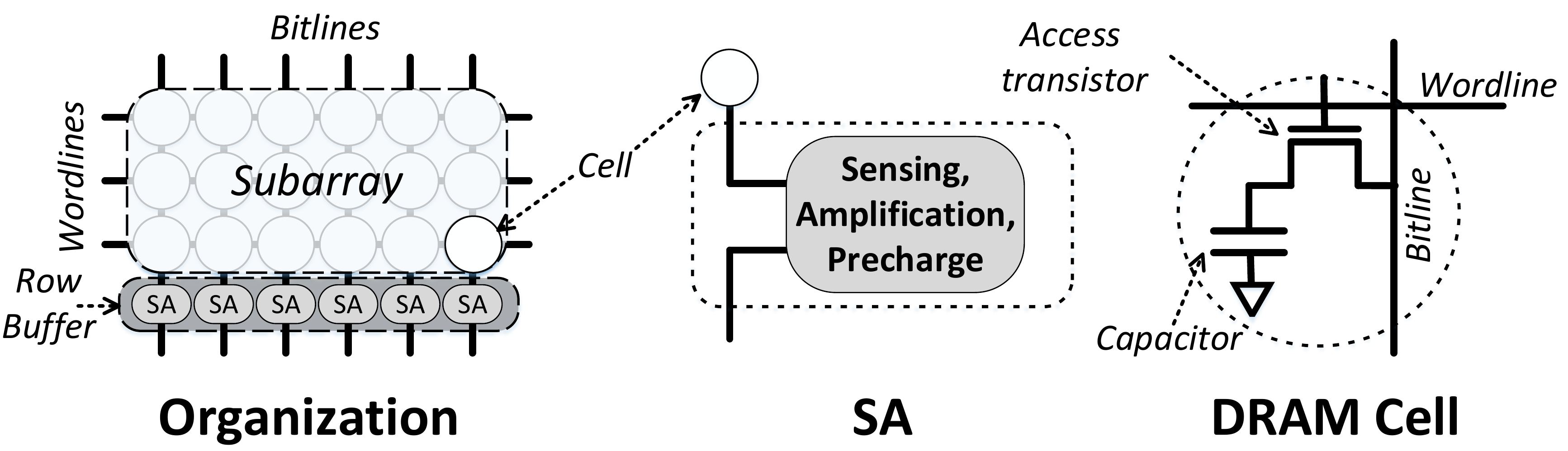}%
    \caption{DRAM organization, sense amplifier, and cell.}
    \label{fig:dram_organization}
\end{figure}

Each DRAM chip consists of multiple \emph{banks}, and each bank contains multiple 2D arrays (or \emph{subarrays}) of DRAM \emph{cells} as shown in Figure~\ref{fig:dram_organization}. The cells are stacked in rows of 4 or 8\,KB that share a \emph{wordline}. Each cell consists of a \emph{capacitor} which stores the data in form of charge, and an \emph{access transistor} controlled by the wordline that connects the cell to the Sense Amplifier through the \emph{bitline}.


\vspace{5pt}\noindent\textbf{DRAM Sense Amplifier (SA)} The Sense Amplifiers (SAs) are used for sensing and amplifying the small charge of the cell capacitor to a CMOS-readable value. A set of SAs connected to a row of cells is called \emph{row buffer}. Figure~\ref{fig:dram_organization} shows how a cell is connected to a \lois{SA} via a \emph{bitline}. The actions related to the functioning of the SA can be summarized into three steps. First, to be able to sense the cell's charge, the SA sets the bitline to the precharge level\ ($V_{dd}$/2). Second, the cell (at $V_{dd}$ or $0$V) shares its charge with the bitline, which produces a small change in the voltage of the bitline ($(V_{dd}/2)\pm \delta$). Third, the SA is activated and it amplifies the delta of the bitline voltage towards the original value of the cell.

\vspace{5pt}\noindent\textbf{DRAM Operation} The memory controller issues three basic commands as part of a DRAM read or write operation: 1) the Activation {\bf  (ACT)}  command senses and amplifies the data from the target row into the row buffer; 2) the read/write ({\bf RD/WR}) command transfers data from/to the row buffer to/from the DRAM bus; 3) the Precharge {\bf  (PRE)} command clears the row buffer and prepares the subarray for subsequent read/write operations (precharges the bitlines).

\begin{figure}[h] \centering
    \includegraphics[width=1.0\linewidth]{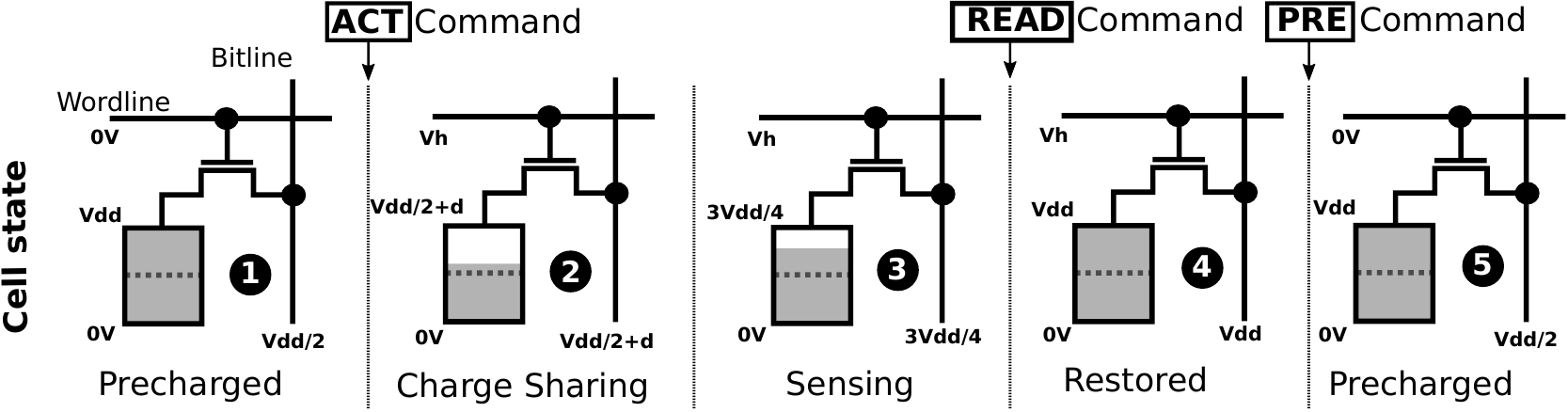}
    \caption{DRAM Activation (ACT), Read (RD) and Precharge (PRE) commands.}
    \vspace{-4mm}
    \label{fig:activation}
\end{figure}

Figure~\ref{fig:activation} details the steps for reading a DRAM cell. \circled{1} Initially, the bitline is precharged to $V_{dd}$/2 with the wordline set to 0V. \circled{2} To access data from DRAM, the memory controller first issues an ACT command, which raises the voltage of the target wordline and connects the cells of that row to the bitline. This causes the deviation of the bitline voltage in one direction (charge sharing). \circled{3} As a result, the sense amplifier senses and amplifies this deviation  (sensing phase). After reaching this phase, the memory controller can issue RD or WR commands. The time needed to finish the ACT command is specified by the timing parameter $tRCD$. \circled{4} The sense amplifier continues to amplify the deviation until the voltage of the cell is fully restored. \circled{5} After that, the controller issues a PRE command to lower the wordline voltage back to 0V and drive the sense amplifier and bitline to $V_{dd}$/2 . The time needed to complete a PRE command is specified by the timing parameter $tRP$. Once precharged, the subarray is ready for the next access.

\section{Overview of \mechanismcap{}}
\label{sec:overview}



\lois{
\mechanism{} is a new \lois{class of low-cost} in-DRAM primitive that enables high-performance implementations of security mechanisms via generating unpredictable yet reproducible values in DRAM. We propose two variants of \mechanism{}, which enable a 1) new class of high-performance and robust PUF, and 2) a new cold boot attack prevention mechanism that does not incur performance and energy overheads at runtime.}

\subsection{\mechanism{} Primitives}
\label{sec:primitive}

\vspace{5pt}\noindent\textbf{\uesa{}  } (\uesaLong{}) \lois{generates unpredictable values in a DRAM SA by exploiting process variation inherent in DRAM SAs}. \lois{We implement \uesa{}} by simply altering the timing of \lois{\emph{two} DRAM signals \minp{during} a standard activation \minp{operation}}. \lois{The generated values can be read by the processor from the SAs without overwriting the data in DRAM, or they can be optionally stored in DRAM.
}

\vspace{5pt}\noindent\textbf{\upla{} } (\uplaLong{}) generates an unpredictable value in \minp{a} DRAM cell by exploiting process variation \minp{inherent in DRAM cells}. \lois{We implement} \upla{} by altering the timing of \minp{only} \lois{\emph{one} DRAM signal \minp{during} a standard precharge \minp{operation}}. \upla{} requires two steps. 
First, \lois{\upla{} sets a DRAM cell to the precharge voltage}. 
\minp{Second, when the cell is next activated, the cell assumes an unpredictable value based on process variation}. \lois{Unlike \uesa{}, \upla{} \lois{requires overwriting} the original content of the cell for generating a value.}

 
 \vspace{5pt}The \mechanism{} primitives \lois{are very similar to DRAM activation and precharge commands}. This makes our approach easy to integrate into commodity DRAM chips, and facilitates its adoption by industry and standards bodies. \lois{Section~\ref{sec:dataplant} describes the circuit-level implementation details of the two primitives, and Section~\ref{sec:eval_lat-ener-area} evaluates their latency, energy, and area.}
\lois{We also propose a new \mechanism{} implementation that generates \emph{deterministic} values by introducing an additional transistor in the SA. Appendix~\ref{sec:deterministic_dataplant} describes and evaluates this implementation.}




\subsection{Implementing Security Mechanism \\Using \mechanism{} Primitives}
\label{sec:overview:mechanisms}


\lois{To demonstrate the potential of \mechanism{}, we use \minp{\upla{} and \uesa{}} to implement \lois{new approaches of two common} security mechanism\minp{s}.} 


\vspace{5pt}\noindent\textbf{\lois{\mechanism{} PUFs}.}
%
\lois{We propose \lois{two} new \mechanism{} DRAM PUFs (based on \uesa{} and \upla{}) that have four unique characteristics. First, the \uesa{}-based PUF does \emph{not} have a destructive read out. Second, the latency of \mechanism{} PUFs is very low (i.e., the same as a precharge or activation latency), which makes \mechanism{} suitable for runtime access. Third, \mechanism{} PUF responses are not as noisy as other state-of-the-art DRAM PUFs, which \minp{results in} more stable PUF responses. Fourth, \mechanism{} PUF responses are very stable across a wide range of temperatures.} 






\vspace{5pt}\noindent\textbf{Preventing Cold Boot Attacks.}
Although DRAM memory is volatile, the stored data does not immediately disappear at power-off. Data can be naturally retained in DRAM cells up to minutes after a power-off~\cite{BAUER2016S65}, which enables \emph{cold boot attacks}~\cite{Halderman2009, Simmons2011, Gruhn2013, Hilgers2014, BAUER2016S65, Yitbarek2017} (i.e., the data can be read as soon as the device is powered back up). 
\lois{We propose a mechanism completely implemented in DRAM chips (i.e., it does not require external DRAM commands) that destroys all the data in DRAM by \lois{automatically} issuing \mechanism{} primitives when the chip is \emph{powered up}}.
Unlike prior mechanisms to prevent cold boot attacks~\cite{Suh:2003, arnold2004ibm, Yang:2005, Duc2006, Rogers2007, Henson:2014}, our \mechanism{}-based mechanism protects against even a computationally \emph{unbounded} adversary, as it makes brute-force attacks impossible.  Our mechanism requires no changes aside from the existence of the \mechanism{} primitives, incurs \emph{no} latency or energy overhead at runtime as it operates only at power-up, \lois{and is secure because it is an automatic mechanism self-contained in DRAM.}



%

\vspace{5pt}Section~\ref{sec:physicallyUnclonable} and Section~\ref{sec:coldbootattacks} discuss the implementation of these mechanisms in detail. \lois{Additionally,} Appendix~\ref{sec:secureDeallocationEval} \lois{describes and evaluates} secure deallocation, an additional security mechanism that can be implemented with \mechanism{}.


\begin{figure*}[t!]
    \centering
    \subfloat[\lois{Activation command}]{
        \centering
        \includegraphics[width=0.31\linewidth]{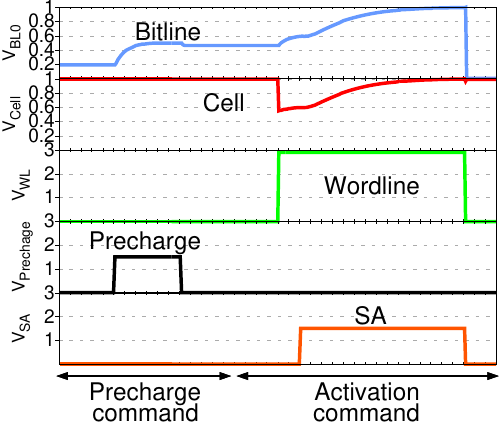}
        \label{fig:regular_act}
    }
    ~
    \subfloat[\uesa{}]{
        \centering
        \includegraphics[width=0.31\linewidth]{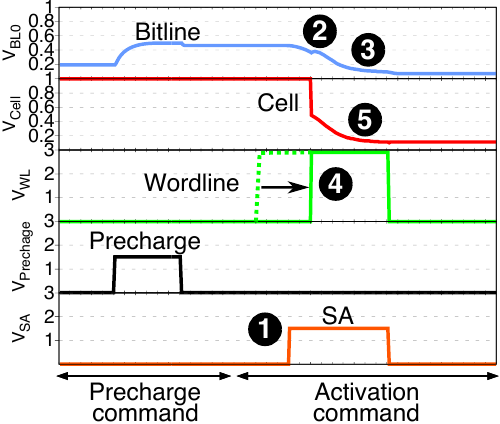}
        \label{fig:uesa_act_timing}
    }
     ~
    \subfloat[\upla{}]{
        \centering
        \includegraphics[width=0.31\linewidth]{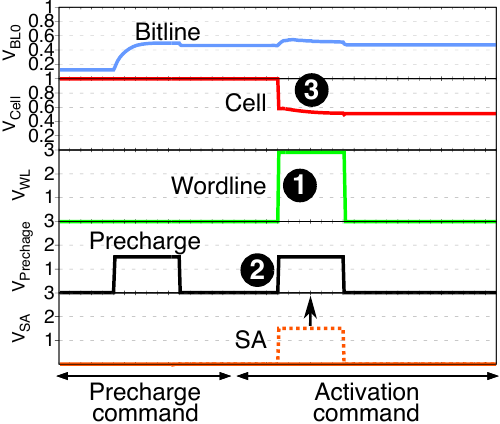}

        \label{fig:upla_act_timing}
    }
    \caption{SPICE simulation of the internal DRAM signals involved in \lois{precharge command} and (a) an \lois{activation command}, (b) a \uesa{} \lois{primitive} (including the optional overwriting of the cell), and (c) a \upla{} \lois{primitive}. In all cases,  $V_{dd}$=1V and the original content of the DRAM cell is ``one" ($V_{Cell}$). \lois{The dashed lines and the arrows ($\rightarrow$, $\uparrow$) illustrate the changes in the timing signals required to implement \uesa{} and \upla{}.}}
    \vspace{-3.5mm}
    \label{fig:timing}
\end{figure*}

\section{\mechanism{} Implementation}
\label{sec:dataplant}

This section shows the implementation details of \lois{two \mechanism variants that have different features. To illustrate how they operate, we use a detailed SPICE model to simulate the SA, cell, bitline and wordline.} \lois{We implement the SA using 55nm DDR3 model parameters~\cite{rambus} and PTM low-power transistor models ~\cite{PTM,zhao2006new}. We use cell/transistor parameters from the Rambus power model~\cite{rambus} (cell capacitance = 22fF; transistor width/height = 55nm/85nm).}



For \lois{reference}, \lois{Figure~\ref{fig:regular_act} shows the simulation of a \lois{standard DRAM activation command}.} 
As explained in Section~\ref{sec:dramorganization}, a \lois{standard activation command 1) raises the voltage of the wordline ($V_{WL}$) to connect the cell to the bitline, which causes a deviation in the bitline voltage, and 2) triggers the SA ($V_{SA}$) for sensing this bitline voltage variation and restoring the charge of the cell towards its original value.}

\subsection{\uesa{} Primitive}
\label{sec:u-esa}

\uesa{} generates unpredictable values by exploiting \lois{mainly SA process variation}. 
\lois{The value generated by \uesa{} has little influence from the process variation of other components (e.g., bitline).}
\lois{The key idea is to trigger the SA \emph{without} raising the voltage of the wordline, i.e., the cell doesn't deviate the voltage of the bitline.}
By doing so, the SA amplifies \lois{the bitline voltage} towards an unpredictable value that 1) doesn't depend on the charge of the cell, \lois{and 2) depends} on the SA process variation. \lois{Once the SA drives the bitline towards the final generated unpredictable value, \uesa{} can optionally write this value into the cell} by \lois{raising} the wordline \lois{voltage}.

Figure~\ref{fig:uesa_act_timing} shows how \lois{\uesa{} generates} a value (\lois{including} the optional overwriting of the cell). \lois{The dashed lines and the arrow ($\rightarrow$) highlight that, when writing to the cell (optional), \uesa{} raises the voltage of the wordline always after triggering the SA logic}. \lois{\uesa{} triggers the SA ($V_{SA}$)} \circled{1}  when the bitline ($V_{BL0}$) is precharged (i.e., $V_{dd}$/2 = 0.5V) \circled{2}, \lois{and the SA} drives the bitline towards a value \lois{(0V in the figure)} that depends on process variation \circled{3}. 
At this point, \lois{the memory controller can issue a read command to get the generated value from the SA.}
\lois{After generating the value, \uesa{} can optionally write the generated value in} the DRAM cell by raising the voltage of the wordline ($V_{WL}$) \circled{4}, \lois{which overwrites the content of the cell with the generated value ($V_{Cell}$)} \circled{5}. \lois{Notice that the voltage of the wordline must be raised after triggering the SA ($\rightarrow$).}

\begin{figure}[h] \centering
    \includegraphics[width=1\linewidth]{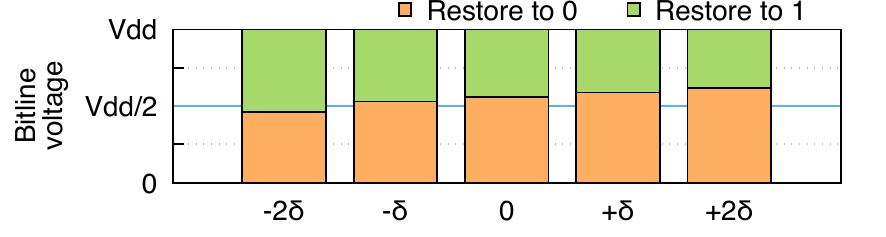}%
    \caption{\lois{Values restored by five SAs with different process variation, depending on the the bitline voltage. \uesa{} generates values always at $V_{dd}/2$ bitline voltage.}}
    \label{fig:process_variation}
\end{figure}

To illustrate the effects of process variation in the values generated by \uesa{}, we perform SPICE simulations for five instances of a common \emph{SA design} \lois{(details in Section~\ref{sec:eval_lat-ener-area})} with small changes in their physical characteristics \lois{that} simulate process variation.  
\lois{Figure~\ref{fig:process_variation} shows the generated value\minp{s} for five SAs examples with different process variation values depending on the voltage of the bitline. \lois{\uesa{} always amplifies a bitline with $V_{dd}/2$ voltage (precharge voltage) \lois{to generate a value}.}
We observe that the \lois{generated} value for \lois{$V_{dd}/2$ bitline voltage} depends on the  SA's process variation: SAs with variation -$\delta$, $0$ +$\delta$ an +2$\delta$ generate a zero value, while -$2\delta$ generates a one value.
These \lois{process} variation at fabrication time cannot be controlled or cloned, and their layout is unpredictable and unique for each device.}


\subsection{\upla{} Primitive}
\label{sec:u-pla}

\upla{} generates unpredictable values by exploiting \lois{mainly DRAM cell process variation}. 
\lois{The value generated by \upla{} has little influence from the process variation of other components (e.g., bitline, wordline, SA).}
\lois{They key idea is to set the cell to the same voltage as the bitline precharge voltage ($V_{dd}/2$) by triggering the precharge logic and raising the wordline voltage at the same time.}
\lois{In the next activation, raising the voltage of the wordline doesn't disturb the bitline voltage (because the cell has the same $V_{dd}/2$ voltage as the bitline), and the SA amplifies towards an unpredictable value. The anomalies and perturbations introduced by the cell process variation~\cite{Lee:2017} are the main factors that determine the unpredictability of the generated value.}

Figure~\ref{fig:upla_act_timing} shows how \upla{} \lois{sets the cell to $V_{dd}/2$ \lois{voltage}}. \lois{The dashed lines and the arrow ($\uparrow$) are for highlighting that \upla{} activates the precharge logic instead of the SA logic}. The wordline \lois{voltage} ($V_{WL}$) \lois{is raised} \circled{1} at the same time \lois{as the precharge logic  is triggered} \circled{2}  ($V_{Precharge}$), which \lois{drives} the cell ($V_{Cell}$) towards $V_{dd}$/2 \circled{3}. 
Our SPICE simulations show that \upla{} consumes the same power independently of the initial value of the cell (as the final value is always $V_{dd}$/2). \lois{Section~\ref{sec:emulating_upla} evaluates the feasibility of \upla{} by emulating the values it generates on real DRAM chips.}

\subsection{\lois{Hybrid-\mechanism{} Primitive}}

\lois{We propose a low-cost Hybrid-\mechanism{} primitive that 1) implements  \uesa{} and \upla{}  into the same DRAM module, and 2) provides a mechanism to select the \mechanism{} implementation at runtime}. 

\lois{Different applications might require different characteristics that a single implementation can not provide}. Our two \mechanism{} implementations have \lois{three main} trade-offs. First, \upla{} relies on destroying the previous content of the cell for generating data, while \uesa{} \lois{can generate data in the SA with or without destroying the DRAM content}. Second, \lois{\upla{} executes faster than \uesa{}, but it requires an additional activation command \minp{to generate the unpredictable value}}. \uesa{}, however, can generate and access the data with only one command.
Third, \upla{} value generation relies on the DRAM cell \lois{process variation}, which are orders of magnitude smaller than the SA. Consequently, \upla{} is potentially  more sensitive to technology scaling effects~\cite{li2010robust}. 

\lois{Hybrid-\mechanism{} primitive enables to choose the primitive that better fits the requirements of the application. Hybrid-\mechanism{} implements \uesa{} and \upla{} together with low hardware overhead by leveraging one free bit in the in-DRAM mode registers (MR) to encode the implementation to use, and the load mode register (LMR) command to change the implementation to use at runtime. In commodity DDR4 modules, the MR3 register has 13 unused bits that enable to select the \mechanism{} primitive independently in 13 different DRAM partitions (e.g., in different DRAM banks)}.

\subsection{\lois{Security and Reliability}}
\label{sec:discussion}
In this section we discuss the security and reliability of \uesa{} and \upla{} primitives.

\vspace{3pt}\noindent\textbf{Security.} \uesa{} does not leak any information about the previous content of the cell because it generates values that are independent of the cell content (Figure~\ref{fig:timing}). \upla{}, as described in Section~\ref{sec:u-pla}, \lois{first discharges the DRAM cells in a row, and it then activates the sense amplifiers.} If an attacker manages to interfere between these two steps, \lois{they} could try to bias the cells towards some particular value before the amplification (e.g., \lois{row hammer}~\cite{Kim2014}). However, these two steps are executed back-to-back in a few nanoseconds, \lois{which is not enough time to induce any row hammering~\cite{Kim2014} or similar attacks that require milliseconds to succeed~\cite{vanderVeen:2016, Aweke:2016:ASP:2872362.2872390}}.


\vspace{3pt}\noindent\textbf{Reliability.} \lois{We observe that the main reliability issue related to DRAM is associated with the reduction of the \minp{amount of charge that can be stored} in a DRAM cell, which can make the \emph{sensing phase} unreliable~\cite{li2010robust,park2015technology}. SAs are in general more resilient to reliability issues~\cite{li2010robust,kim2015architectural} because \minp{they} are three orders of magnitude larger than DRAM cells~\cite{rambus}. Based on these observations, we make two conclusions. First, we do not expect major reliability issues on \uesa{} as it does not use the cells for generating values. Second, although \upla{} uses the DRAM cells for generating values, \upla{} does not have a sensing phase because the cell is set to $V_{dd}/2$ voltage. Therefore, we expect the reliability of \lois{a US/UC-\mechanism{} primitive} to be higher than the reliability of a regular \lois{activation/precharge}.} 


\subsection{\lois{Hardware Cost}}
\label{sec:dataplant_hw_cost}
The hardware cost of implementing \mechanism{} in DRAM is very low. 
%
%
\lois{Incorporating our new \mechanism{} primitives require very few  modifications in the control logic that generates the control signals. The changes are limited to add a few extra logic gates to delay the wordline signal in a regular activation (\uesa{}), and to trigger the access transistor and the precharge logic at the same time (\upla{}). To the best of our knowledge, there is no public information about how vendors implement the control logic, or what is the specific circuit design of that logic (see Appendix~\ref{sec:DRAM_logic}). The hardware cost of our primitives is very low in any case, because our mechanism can reuse most of the logic for generating the activation and precharge timing signals.
} 

\section{\mechanism{} PUFs}
\label{sec:physicallyUnclonable}

\lois{ A PUF is a hardware primitive that maps a set of challenges to a set of static random responses that are derived from the physical characteristics of \lois{an} integrated circuit (e.g., process variation). A PUF can be used as a building block for implementing low-cost authentication protocols~\cite{Majzoobi2012, Hammouri2008, Rostami2014, Che2015} and key generation applications~\cite{Roel2012,Yu2012,Paral2011}.}

\lois{One or more parameters (e.g., the address of a memory segment, temperature, etc.)  define a challenge, and the data read from DRAM is the response to that challenge. Together, they define a Challenge-Response pair (CR pair). In this work, we use the address of the segment as the only parameter that defines a challenge.}



\vspace{5pt}\noindent\textbf{Limitations of State-of-the-art DRAM PUFs.} 
\label{sec:stateoftheart_PUFs}
Prior DRAM-based PUF proposals exploit variations in DRAM start-up values~\cite{Tehranipoor:2015}, DRAM write access latencies~\cite{Hashemian:2015}, DRAM cell retention failures~\cite{Sutar2016, Xiong2016, Keller2014} and reduced DRAM timing parameters~\cite{kim2018,talukder2019prelatpuf}. There are \lois{five} main \lois{limitations} with \lois{most} of these approaches. First, \lois{most} of these PUFs rely on the charge that is contained in DRAM cells, thus all the content of the memory region employed for the PUF is irreversibly overwritten. Using these PUFs \lois{requires} either 1) exclusive memory regions for PUFs or 2) copying and restoring the original contents for each PUF challenge request. Second, most DRAM PUFs~\cite{kim2018} have high evaluation time, \lois{which can potentially  cause system interference when the PUF is accessed at runtime}. \lois{Third, many of these PUFs require heavy filtering mechanisms to deal with the inherent noisy nature of the DRAM responses, which increases the evaluation latency and the reliability of the responses. Fourth, the responses to the same challenge suffer great variations with temperature changes, which is an issue in systems with a non-controlled environment (e.g., IoT devices in the wild). Fifth, some DRAM PUFs are data dependent, which might cause mismatching responses that depend on the content of the memory.} 
\lois{Our Hybrid-\mechanism{} PUFs overcome all of these \lois{five} limitations.}

\vspace{5pt}\noindent\textbf{\UESAPUF{}.} \lois{\UESAPUF{} has six key distinctive properties. 
First, \UESAPUF{} does not require \minp{destroying} the current memory content, because it does not need to assert the wordline to generate a response.
Second, \UESAPUF{} responses do not depend on the actual content of the cell, because the charge of the cell shared and  amplified. 
Third, \UESAPUF{} can be evaluated with a latency as low as the latency of a regular activation. 
Fourth, a \UESAPUF{}  response is less noisy than most state-of-the-art DRAM PUFs, which \minp{enables} lighter filtering mechanisms and reduced latency. 
Fifth, \UESAPUF{} is particularly robust because SAs are much less sensitive to environmental conditions, interference from other elements, and scaling issues~\cite{Kim2014}. 
Sixth, the downside of \UESAPUF{} is that it has a small CR pair space that is limited by the number row buffers in DRAM ($\sim$8MB in a 4GB DRAM).}



\vspace{5pt}\noindent\textbf{\UPLAPUF{}.} \lois{ \UPLAPUF{} has six distinctive properties. 
First, \upla{} has a CR pair space as large as the DRAM capacity, because it uses DRAM cells to generate responses.
Second, the evaluation latency of \UPLAPUF{} is as low as the latency of a precharge operation.
Third, \UPLAPUF{} is more reliable under temperature variations than the state-of-the-art solutions, as we demonstrate in our evaluation (Section~\ref{sec:eval_lat-ener-area}).
Fourth, \UPLAPUF{} requires an additional activation to read out the PUF response.
Fifth, \UPLAPUF{} is more sensitive to scaling issues than \UESAPUF{}, because the generated responses depend mainly on the DRAM cells.  
Sixth, \UPLAPUF{} necesarily \minp{destroyes} the current memory content. 
}

\vspace{5pt}\noindent\textbf{Hybrid-Dataplant PUF.} \lois{To get the best properties of our two \mechanism{} PUFs, we can easily implement \UESAPUF{} and \UPLAPUF{} in the same DRAM module with minimal hardware overhead. The application can  choose, according to its requirements, the PUF to use (\uesa{} or \upla{}) by configuring the MRs accordingly (see Section~\ref{sec:discussion}).}


\label{sec:isasupport}
\vspace{5pt}\noindent\textbf{System-Level Support for Accessing \mechanism{} PUFs.} There are \lois{at least} two ways of enabling software access to \lois{\mechanism{}} PUFs, either by adding a new instruction to the instruction set architecture for reading the PUF \lois{response}~\cite{seshadri2013,Chang2016LISA}, or by using a dedicated address range to map the PUF operations to regular load instructions.

\lois{On the DRAM side, there are two implementation options. First, adding a new mode that provides the \uesa{} and \upla{} functionalities instead of the regular activation and precharge commands. The mode can be selected by changing dedicated MRs (see Section~\ref{sec:discussion}). Second, introducing a new command in the DDR specification. The new command has the same general requirements as a regular activation (\uesa{}) or as a regular precharge (\upla{}). We can integrate the new command in the JEDEC standard specification~\cite{jedec2012jedec} without extra cost, as there is unused, reserved space as part of the standard for new commands.}

\vspace{5pt}\noindent\textbf{Security Analysis} 
%
\lois{\mechanism{} PUFs \minp{are} more secure, less noisy, and more stable with temperature changes than state-of-the-art DRAM PUFs. The \mechanism{} PUF responses responses pass all the NIST randomness tests. \minp{Note} that all DRAM PUFs \minp{have} a CR pair space that is limited by the DRAM capacity. Security applications that use DRAM PUFs as a building block have to engineer the security mechanism carefully to avoid, for example, that an attacker with physical access to the device characterizes the \minp{entire} device and compromise its security.}

\section{Preventing Cold Boot Attacks}
\label{sec:coldbootattacks}

This section shows how \mechanism{} enables an efficient and simple mechanism to prevent cold boot attacks. Cold boot attacks~\cite{Yitbarek2017,Halderman2009,Simmons2011,Gruhn2013} are possible because the data stored in DRAM is not immediately lost when the chip is powered-off. This is due to the capacitive nature of DRAM cells that can hold their data up to some seconds~\cite{BAUER2016S65}. This reminiscent effect can be even more significant if the DRAM module is cooled down. Taking advantage of this property, an attacker can either take the victim's DRAM module off and place it in a system under \lois{their} control with minimal information loss, or  boot a small special purpose program from a cold reset to recover the secret information. 

\subsection{State-of-the-Art Defenses}
\label{sec:state-of-the-art_defenses}

\lois{There are three classes of mechanisms for preventing cold boot attacks.} 
First, mechanisms that rely on encrypting memory either explicitly~\cite{Suh:2003, arnold2004ibm, Yang:2005, Duc2006, Rogers2007, Henson:2014, amd_sme, Yitbarek2017}, or implicitly through some CPU extensions (e.g. Intel SGX~\cite{costan2016intel}). These mechanisms are effective and secure, but are too complex and expensive \lois{(in terms of energy and performance overhead)} to be implemented in many low-cost devices. 
\lois{Second, modern systems scramble the data in the memory controllers, which helps to obscure the DRAM contents. This mechanism is simple and \minp{is also required} for other purposes (e.g., improve signal integrity in the DRAM bus), but \minp{it has} been shown to be insecure against cold boot attacks~\cite{Yitbarek2017}.}
Third, the mechanism proposed by the Trusted Computing Group (TCG)~\cite{computing2009tcg} to reset the DRAM content upon power-off (or power-on if the last power-off was not clean). This mechanism is implemented on the host platform firmware and depend on the OS, which makes it vulnerable to attacks~\cite{fsecureattack}. 

\subsection{Threat Model}
\label{sec:threatModel}
We tackle an attacker that gains physical access to a live uncompromised machine/device for an unlimited amount of time and whose goal is to obtain some information stored in the device's DRAM. Note that, for any interesting information to be present in DRAM the attacker should get the memory while it is still \lois{powered-on}. We then assume that, as part of the attack, the DRAM chip is powered-off for an arbitrarily short amount of time. This power loss occurs when transplanting the DRAM module to an attacker-controlled machine and during attacks that reboot the victim machine to load a malicious OS. Note that some computers allow warm reboots, in which the power is not cut off. Our cold boot attack prevention mechanism is not compatible with \lois{those systems}.

We are not aware of alternative methods to power-on the DRAM, other than using the corresponding DRAM PINs. Therefore, until a technique that can attach a stable external power supply (while the chip is already powered on) to the DRAM chip is engineered, transplanting a chip from one machine to another would inevitably involve a power loss. However, even if this is possible, we speculate that it would require expensive specialized equipment, significantly increasing the cost of an otherwise cheap attack.

In summary, to the best of our knowledge, transplanting the DRAM and rebooting to a different OS are the only ways to perform a \lois{cold boot attack} today, and both involve a power loss. In particular, we are not aware of any other techniques that allow measuring the charge in the DRAM capacitors, including x-ray techniques.

\subsection{Destroying Data at \lois{Power-On}}

\lois{We make the observation that it is possible to protect from cold boot attacks by deleting all memory contents during the DRAM power-on.} 
Based on this observation, we propose two new \lois{cold boot attack} prevention mechanisms. First, \emph{Self-destruction}, a low-cost \lois{in-DRAM} mechanism based on \mechanism{} that destroys all the DRAM content without the intervention of the memory controller. Second, \emph{Command-based Destruction}, a low-cost mechanism orchestrated by the memory controller that allows a more flexible implementation at the cost of providing weaker security guarantees.

\subsubsection{Self-destruction}
\label{sec:self-destruction}

\lois{The key idea of \lois{Self-destruction} is to refresh the whole DRAM memory in self-refresh (SR) mode at power-on, but \lois{using} \mechanism{} primitives instead of activation commands.} This way, the DRAM chip executes a \emph{destructive} \lois{DRAM refresh that} can be performed autonomously without the intervention of the memory controller.

The basic principle of a DRAM refresh is to \lois{perform} an activation and a precharge command to the row to be refreshed. As we show in Section~\ref{sec:dataplant}, \uesa{} is very similar to an activation command, and \upla{} is very similar to a precharge command, which allows to easily incorporate them in the refresh operations, leveraging the circuitry that launches regular SR cycles. With Self-destruction, the data is destroyed in a complete SR window, i.e., 64ms (32ms for LPDDR). During the destructive SR, the DRAM does not allow any memory commands  to ensure the atomicity of the process.

\vspace{5pt}\noindent\textbf{Self-destruction in a Burst refresh.} \lois{Burst} refresh is a refresh mode that is available on Low-power \minp{DDRn (LPDDRn)} devices. The main idea of the burst refresh is to complete all the required refreshes in a single burst, with the goal of meeting the deadlines of real-time applications. Our Self-destruction mechanism is also compatible with this refresh mode that allows destroying data much quicker \lois{(e.g., 9ms for a 4GB DDR4 memory module)}.




\vspace{5pt}\noindent\textbf{Security Analysis.}
\lois{Our mechanism is automatically triggered in DRAM when power is detected, without requiring external actions}. Therefore, the security of Self-destruction depends on the reliability of the power-on detection circuit of the DRAM module. There are two ways in which an attacker can potentially bypass this circuit. We describe \lois{both ways} and explain why, in practice, they do not pose a security threat. 

First, an attacker could operate DRAM at low voltage on the compromised system using, for instance, Dynamic Voltage and Frequency Scaling (DVFS), with the goal of \lois{\emph{not}} triggering the power-on detection circuit. The power-on circuit triggers when it detects a voltage ramp up from $0V$, but it does not need to reach $V_{dd}$ (it triggers as long as a voltage ramp up starting from $0V$ is detected). Therefore, operating the DRAM at very low voltage would not help the attacker.\footnote{The attacker might try to operate the device at a voltage close to $0V$ such that the power-on circuit cannot detect the ramp up, however, at such low voltage the DRAM would not be operational.}

Second, an attacker could fry the \lois{DRAM power-on} detection mechanism. In practice, however, the FSM that initializes the chip is in the same internal controller that regulates other functions (i.e., timing \lois{signals} for activate, precharge, and other commands). Consequently, frying that component would most likely make the whole DRAM unusable.

\vspace{5pt}\noindent\textbf{Hardware Cost Analysis.}
The hardware cost of implementing our Self-destruction mechanism in DRAM is very low. The implementation of \mechanism{} has very low overhead (Section~\ref{sec:discussion}), and the logic to trigger a self-refresh window at \lois{DRAM power-on} is negligible. Triggering \mechanism{} instead of regular activations in the refresh process requires minimal modifications on the in-DRAM \lois{control logic (see Section~\ref{sec:dataplant_hw_cost})}.

\subsubsection{Command-Based Destruction.} 

The key idea of Command-Based Destruction is to force DRAM to \lois{obey} a particular sequence of commands \lois{issued} from the memory controller that leads to the \lois{destruction} of the whole memory content during the initialization procedure. The mechanism can be implemented with regular write commands, with Rowclone~\cite{seshadri2013}, with Lisa-clone~\cite{Chang2016LISA} or with our \lois{new} \mechanism{} primitives. 

The Command-Based Destruction relaxes the security guarantees since it is conducted by the memory controller. An attacker could easily bypass this procedure by using a customized memory controller or a programmable one~\cite{hassan2017}. \lois{To solve this issue, we add} a mechanism in DRAM that ensures the execution of the appropriate sequence of commands in the initialization procedure. \lois{The mechanism uses a latch that} indicates when the DRAM is performing the initialization and, during this phase, \lois{the DRAM chip} filters out any other command. Implementing this \lois{mechanism} requires \lois{adding a new} FSM in DRAM, which adds hardware and energy overhead to the existing circuitry. Also, a DRAM module that implements Command-Based Destruction \lois{can operate only} with compatible memory controllers \lois{(i.e., no backward compatible)}.


\vspace{5pt}\noindent\textbf{Security Analysis.}
Compared to Self-destruction, Command-based Destruction has weaker security guarantees, because it is not self-contained in-DRAM, and it does not destroy the memory contents automatically \lois{in-DRAM} at \lois{power-on}. Nevertheless, it is very challenging to bypass the DRAM FSM that disables read commands until the memory is destroyed by the memory controller. 
Compared to TCG \lois{(Section~\ref{sec:state-of-the-art_defenses})}, Command-based Destruction provides better security guarantees, as our mechanism does not provide any software interface to control the DRAM initialization mechanism. 

\vspace{5pt}\noindent\textbf{Hardware Cost Analysis.}

\lois{Self-destruction does not require modifications in the memory controller, and it requires minimal changes in the DRAM logic that controls the signals to issue \mechanism{} commands instead of regular activation and precharge commands.}

Compared to Self-destruction, Command-based Destruction is more complex to integrate into current systems. It requires the modification of the memory controller, and it requires the addition of dedicated DRAM logic to ensure \lois{the integrity and atomicity of the destruction protocol}. 

\lois{Command-based destruction issues} Rowclone/Lisa/\mechanism{} requests from the memory controller. Similar to the PUF mechanism (Section~\ref{sec:physicallyUnclonable}), there are two options to issue these requests. First, \lois{adding a new DRAM command to the DDR JEDEC standard specification~\cite{jedec2012jedec} by leveraging the unused and reserved bits available in the standard protocol. Second, adding new logic in the in-DRAM command decoder that decodes existing commands (e.g., activation, precharge) into Rowclone/Lisa/\mechanism{} commands depending on some new configuration bits in the SR registers  (similar to Section~\ref{sec:discussion})}. 
\section{Evaluations}
\label{sec:evaluation}

We evaluate the \mechanism{} \lois{primitives} (Section~\ref{sec:eval_lat-ener-area}),  the \mechanism{} DRAM PUFs (Section~\ref{sec:eval_randomnes}) and our cold boot attack prevention mechanism (Section~\ref{sec:eval_coldBootAttacks}).

\subsection{\mechanism{}: Latency, Energy, and Area}
\label{sec:eval_lat-ener-area}

\vspace{3pt}\noindent\textbf{Methodology.} 
\lois{In this section we study the latency and energy overhead of \mechanism{} primitives for generating and overwriting values in a single DRAM row. We compare our \mechanism{} primitives to the state-of-the-art mechanisms for \emph{copying} data within DRAM, namely Lisa-clone~\cite{Chang2016LISA} and Rowclone~\cite{seshadri2013}. Rowclone and Lisa-clone propose in-DRAM methods to initialize data to zero by copying a reserved row filled with zeros to the destination row. Both solutions modify the internal architecture of DRAM and slightly reduce the DRAM's capacity since they need helper data to work.} \lois{We \lois{compare \mechanism{} primitive against Rowclone and Lisa-clone} because there are no other works that \emph{generate} data within DRAM in the same way \mechanism{} does.} 

We estimate the latency of \lois{\uesa{}, \upla{}, Rowclone and Lisa} assuming DDR3 timing constraints. We calculate \lois{their} energy consumption by using the activation and precharge energy consumption described in the power model of the DRAMPower simulator~\cite{chandrasekar2012drampower}. 


\vspace{3pt}\noindent\textbf{Latency and Energy Results.}
Table~\ref{table:energy_latency} shows the absolute \lois{value} and the reduction of latency and energy of the evaluated techniques, when \lois{generating a value in a} 8\,KB DRAM row. \lois{We also show the in-DRAM latency and energy consumption of standard activation and precharge commands.}  \lois{The baseline generates data by overwriting the memory contents with regular write commands from the memory controller.} We make \lois{three} major observations. First, the latency and energy consumption of our two \mechanism{} primitives are significantly reduced compared to the baseline, Lisa-clone and Rowclone. Second, \upla{} is significantly faster than \uesa{}, mainly because it avoids the activation of the SA. \lois{However, the \upla{} numbers on the table do not include the additional activation command needed read the values out of DRAM (see Section~\ref{sec:u-pla}). Third, the latency and energy consumption of \uesa{}/\upla{} is the same as a standard activation/precharge command.}

\begin{table}[h]
    \caption{Latency and energy of different in-DRAM primitives for overwriting 8KB of data, and standard activation and precharge commands.} 
    \centering
    \begin{threeparttable}
    \footnotesize
    \begin{tabular}{| r | c | c | c | c | c |}
      \multicolumn{1}{c}{} & \multicolumn{2}{c}{ {\bf Absolute}} & \multicolumn{2}{c}{{\bf Reduction}}\\
      \hline
         {\bf Primitive } & {\bf Lat. (ns)} & {\bf Ener. (nJ)} & {\bf Lat.} & {\bf Ener.}\\
        \hline
        \hline
        Baseline  & 546 & 2000 & 1.0x & 1.0x \\
        \hline
        Lisa-clone  & 148.5 & 90 & 3.67x & 22.2x \\
        \hline
        Rowclone  & 90 & 50 & 6.06x & 41.5x \\
        \hline
        \hline
        \lois{Activation} & 35 & 17.3 & 15.6x & 116x\\
        \hline
        \lois{Precharge} & 13 & 17.2 & 42x & 116x\\
        \hline
        \hline
        \textbf{\uesaShort{}} & 35 & 17.3 = 7.3 + 10 & 15.6x & 116x \\
        \hline
        \textbf{\uplaShort{}} & 13 & 17.2 & 42x & 116x\\
        \hline
    \end{tabular}
    \end{threeparttable}

    \label{table:energy_latency}
\end{table}


Table~\ref{table:energy_latency} also shows the \mechanism{} energy breakdown (value generation + overwriting). The energy consumption is very \lois{close} on the two \lois{\mechanism{}} implementations because of two main reasons. First, the two implementations need to route the address within DRAM, which is one of the main sources of energy consumption (around 40\%). Second, the energy consumption of the sense amplifier (used in \uesa{}) and the precharge logic (used in \upla{}) are similar (around 40\%). Notice that overwriting in \uesa{} is optional, hence they require only 7.3nJ and 8nJ respectively to generate an 8KB value, while in \upla{} both processes are indivisible, requiring always 17.2nJ for generation+overwriting.


\lois{Our SPICE simulations show that the power demanded by \uesa{} can vary up to 5\% depending on the initial value contained on the cell.}

\vspace{3pt}\noindent\textbf{Area Overhead.} 
%
\uesa{} and \upla{} has \lois{negligible} area overhead caused by the additional logic that controls the signal timings \lois{(more detail in Section~\ref{sec:dataplant_hw_cost})}.
Lisa-clone has an area overhead of 1\% caused by the additional isolation transistors, additional control logic, and one additional zero-filled row per bank. The overhead of Rowclone (0.2\%) is caused by the additional zero-filled row per subarray.

\subsection{Evaluating the Quality of \mechanism{} PUFs}
\label{sec:eval_randomnes}


\lois{To evaluate \uesa{} and \upla{} PUFs, we reproduce the responses of \uesa{} with SPICE simulations, and we reproduce the responses of \upla{} in \emph{real} DRAM chips with an FPGA-based infrastructure.}

\subsubsection{Simulating \UESAPUF{} with SPICE}
\label{sec:simulating_uesa}
We evaluate \UESAPUF{} with SPICE simulations. \lois{Unfortunately,} it is unfeasible to conduct experiments on real DRAM chips, as \uesa{} requires \minp{changes to the internal DRAM timings, which are hard-wired in commodity DRAM chips}.

\vspace{3pt}\noindent\textbf{Methodology.} To show the effects of process variation on the values generated by \uesa{}, we evaluate a detailed SA SPICE model \lois{(see Section~\ref{sec:dataplant})} using Monte Carlo simulations. 
We model variations in all the affected components of the \lois{SAs} (transistor length/width/threshold voltage). Our SA model always generates `1' bits in absence of process variation. When we introduce process variation into the simulation, we observe that some SAs generate `0' bits as well (\lois{called} \emph{unpredictable values}). We run 100,000 simulations for each variation.

\vspace{3pt}\noindent\textbf{Results.}
Table~\ref{table:common_design} shows the percentage of SAs that generate unpredictable values for different levels of process variation and different temperatures.

\begin{table}[h]
    \caption{Effect of \lois{Process Variation (PV)} and temperature on the unpredictability of the values generated by  \uesa{}.}
    \centering
    \footnotesize{}
    \setlength\tabcolsep{3pt} 
    \begin{tabular}{| c | c | c | c | c || c | c | c | c |  }
      
        \multicolumn{1}{c}{ } & \multicolumn{4}{c}{ {\bf PV effects}} & \multicolumn{4}{c}{{\bf Temperature effects \lois{(4\% PV)}}} \\
        \hline
         & {\bf 2\%} & {\bf 3\%} & {\bf 4\%} & {\bf 5\%} &  {\bf 30$^\circ$C} & {\bf 60$^\circ$C} & {\bf 70$^\circ$C} & {\bf 85$^\circ$C}  \\
        \hline
         {\bf Unpred.} & 0\% & 0\% & 0.02\% & 0.19\%  & 0.02\% & 0.19\% & 0.21\%  & 0.15\% \\
        \hline
    \end{tabular}
    \label{table:common_design}
\end{table}

We make two main observations. First, small process variations ($<$4\%) are not enough to generate unpredictable values. Second, large process variations increase the unpredictability of the generated values. As the technology scales, process variation becomes more significant, which increases the unpredictability of the values generated by \UESAPUF{} (i.e., \lois{it increases} the PUF quality).  Third, temperature changes do not cause significant variation in the unpredictability of the generated values. 

\subsubsection{Evaluating \UPLAPUF{} Responses in Real DRAM Chips}
\label{sec:emulating_upla}

\lois{
We evaluate the quality of the \UPLAPUF{} responses with a new methodology that allows us to recreate the functional behavior of the  \UPLAPUF{} without implementing it in real DRAM chips. Notice that it is not possible to implement \UPLAPUF{} in commodity DRAM devices, because \mechanism{} requires changes to the internal DRAM timings. We perform an exhaustive evaluation using \lois{136} \emph{real} DRAM chips from 15 modules.   
}

\vspace{3pt}\noindent\textbf{Methodology.}
\lois{An \upla{} primitive 1) sets a cell to $V_{dd}/2$ with the precharge logic, and 2) activates the SA to generate an unpredictable value from that cell. As we don't have the resources to make a real DRAM implementation, we emulate this behavior in \emph{real} DRAM chips in two steps. First, based on the observation that a DRAM cell leak towards $V_{dd}/2$, we disable the DRAM refresh for 48 hours with the goal setting the cell to $V_{dd}/2$. Second, we activate this cell to obtain the PUF response. This methodology allows us to reproduce the responses that would produce a real \UPLAPUF{} implementation. Recall that discharging the cells would take a few nanoseconds (\emph{not} 48h) in a real implementation (Section~\ref{sec:eval_lat-ener-area}). We perform our experiments with a customized memory controller built with SoftMC~\cite{hassan2017} and a Xilinx ML605 FPGA on 136 different DDR3 DRAM chips from three major vendors.}

\lois{Table~\ref{table:dram_chips} shows the main characteristics of the 136 DRAM chips we use in our evaluation, including vendor, number of chips (\#Chips), capacity of a chip (capacity/chip), frequency, and voltage.}

\begin{table}[h]
    \caption{\lois{Characteristics of the 136 DDR3 DRAM chips used in our evaluation.}}
    \centering
    \footnotesize{}
    \setlength\tabcolsep{3pt} 
    \begin{tabular}{| c | c | c | c | c |  }
        \hline
        {\bf Vendor} & {\bf \#Chips} & {\bf capacity/chip} & {\bf Frequency} & {\bf Voltage}  \\
        \hline
        A & 32 &  512MB & 1600 & 1.35V (DDR3L) \\ 
        \hline
        A & 32 &  512MB & 1600 & 1.5V (DDR3) \\ 
        \hline
        B & 32 &  256MB & 1333 & 1.5V (DDR3) \\ 
        \hline
        B & 8 &  512MB & 1600 & 1.35V (DDR3L) \\ 
        \hline
        C & 32 &  512MB & 1600 & 1.35V (DDR3L) \\ 
        \hline
    \end{tabular}
    \label{table:dram_chips}
\end{table}

\lois{Emulating the \upla{} PUF functional behavior with our methodology is challenging,} because DRAM cells can retain their content for a long time~\cite{Liu2013}, i.e., not refreshing the DRAM does not guarantee that a cell will be \lois{end up with the precharge voltage ($V_{dd}/2$)}, even after a long period.
 
To deal with this issue, we tailor a custom test to determine if a cell is \lois{set to the precharge voltage}. As discussed in Sections~\ref{sec:u-esa} and~\ref{sec:u-pla}, when a cell is \lois{set to the precharge voltage}, the value that \lois{\upla{} generates}  should be always the same regardless of the initial value of the cell. Based on this observation, our test analyzes the final value of a DRAM cell after 48 hours without refresh, \lois{in} two different scenarios:
1) all initial values are zero and 2) all initial values are one. 
The test has two possible outcomes.
First, the \emph{test passes} if the final value is the same regardless of the initial value. Thus, we can conclude that the cell \lois{is set to the precharge voltage}. In this case, the final value should be the one that a real \upla{} implementation would generate. Second, the \emph{test fails} if the final value is different. In that case, we cannot conclude that the cell is set to the precharge voltage (i.e., we cannot infer the value generated by \upla{}), so we do not consider that cell in our \lois{results}.

\vspace{3pt}\noindent\textbf{Results.}
Our experiments cover between 34\% to 99\% \lois{of all cells}, which \lois{are the} cells that \lois{end up with the precharge voltage using} our methodology. The percentage of generated values that are unpredictable because of process variation is between 0.01\% and 0.22\%, \lois{which is in line with the results we obtained with our SPICE simualtions}\footnote{\label{foot_entropy}We evaluate the randomness of the values generated by UC-Dataplant in Section~\ref{sec:nist}}. 
To measure the \emph{uniqueness} and \emph{similarity} of a PUF, we apply Jaccard indices~\cite{jaccard1901etude} as suggested by prior works~\cite{schaller:2017, Xiong2016, aysu2017new, kim2018}. We determine the Jaccard indices by taking two sets of unpredictable values ($u1,u2$), i.e, two sets of PUF responses, from two memory segments, and calculating the ratio of their shared values over the full set of unique unpredictable values \footnotesize $\dfrac{|u1 \cap u2|} {|u1\cup u2|}$\normalsize. A ratio close to 1 represents high similarity, and  a ratio close to 0 represents uniqueness.

We use the term Intra-Jaccard for representing the similarity of two sets from the \emph{same} memory segment, and Inter-Jaccard for representing the uniqueness of two sets from \emph{different} memory segments. An ideal PUF should have an Intra-Jaccard index close to 1 (a unique challenge has a unique response), and an Inter-Jaccard index close to 0 (different challenges have different and random responses).

We compute the distribution of Intra- and Inter-Jaccard indices obtained by running experiments on 
\lois{136} different DRAM chips with segments of 8KB (\lois{this size is used by prior work~\cite{kim2018}}). We calculate the Intra-Jaccard indices for 10,000 random pairs of memory segments (each pair composed of two responses from the same memory segment), and the Inter-Jaccard indices for 10,000 random pairs of memory segments (each pair composed of two responses from different memory segments) from all \lois{DRAM chips}.

We compare \UPLAPUF{} with the DRAM latency PUF~\cite{kim2018}. The DRAM Latency PUF accesses DRAM with reduced timing parameters, which causes some read failures that \lois{fulfill the requirements of a} good PUF. We \lois{implement} the DRAM Latency PUF \lois{by reducing $tRCD$ to 2.5ns}, as it is the timing \lois{value} that reports the best results \lois{in} our setup. For improving the repeatability of the responses, the DRAM latency PUF implements a filtering mechanism that removes the cells with low failure probability from the PUF response. To this end, the mechanism reads the memory segment 100 times, and it composes a response that contains only the failures that repeat more than 90 times~\cite{kim2018}. 

\lois{The values generated by \upla{} are much less noisy than the values obtained by reducing the access latency, so \UPLAPUF{} needs a much more lightweight filtering mechanism: \lois{we actually observe that one read is enough to get a robust \UPLAPUF{} response in most cases, but we apply a conservative filter of 10 \upla{} requests to obtain more robust PUF responses in worst case conditions}. While a DRAM latency PUF with a lightweight filtering mechanism \lois{(e.g., 1-10 reads)} could be as fast as Dataplant PUFs, the PUF quality \lois{would decrease significantly} (Section~\ref{sec:puf_evaluation_time}), compromising the functionality and security of \lois{the PUF-based authentication protocol}.}

\begin{figure}[h] \centering
    \includegraphics[width=1.0\linewidth]{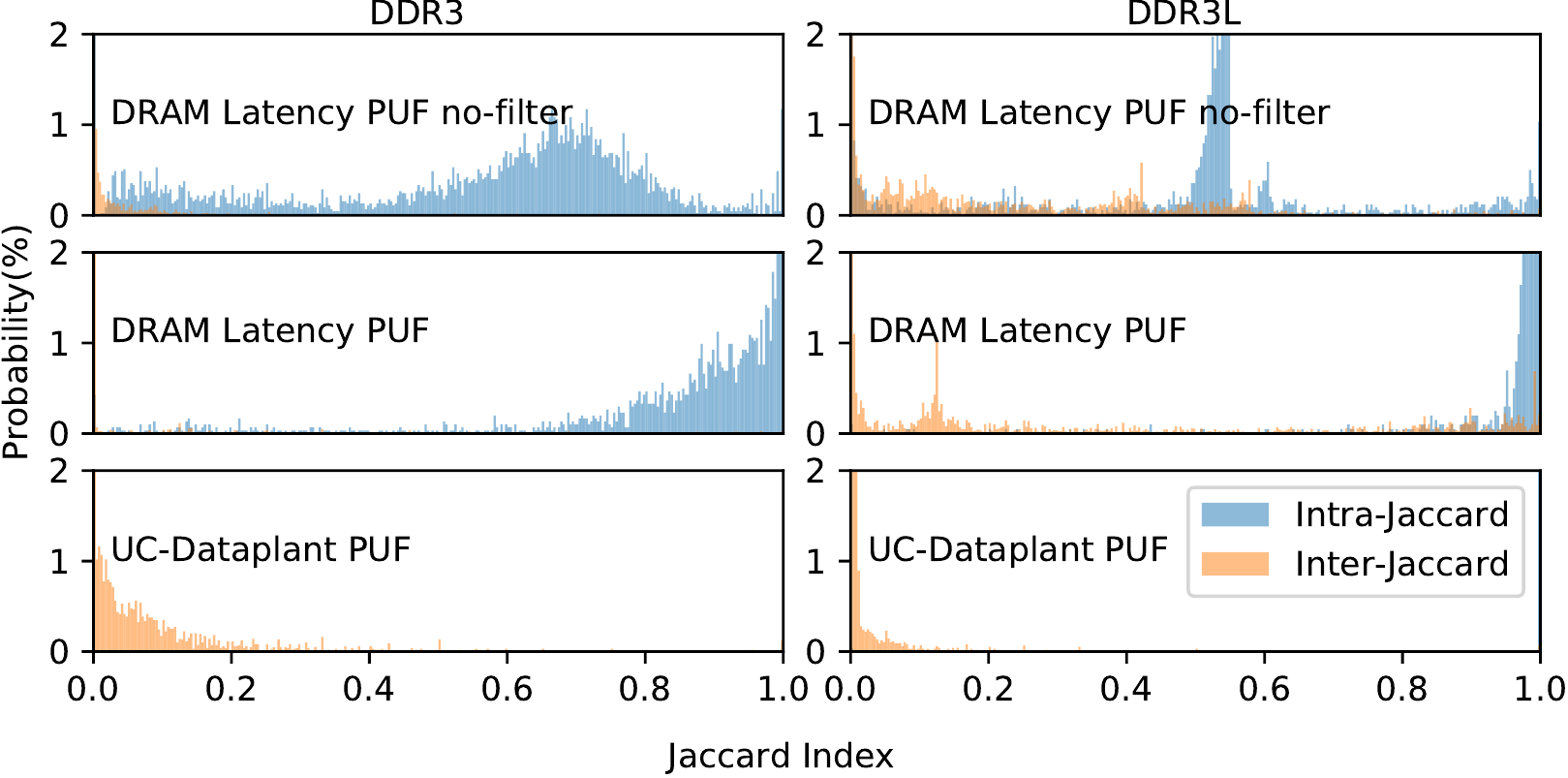}
    \caption{Jaccard indices obtained with the DRAM latency PUF (with and without filter), and with the \UPLAPUF{}, on both DDR3 and DDR3L \lois{chips}.}
    \label{fig:jaccard}
\end{figure}

Figure~\ref{fig:jaccard} shows the Intra- and Inter-Jaccard indices of \UPLAPUF{} and DRAM Latency PUF~\cite{kim2018} for \lois{64 DDR3 chips} \lois{operating at 1.5V} and \lois{72 DDR3L chips} \lois{operating at 1.35V}. 

We make four main observations. 
First, the \UPLAPUF{} shows very good Intra-Jaccard indices (almost all indices are one), and pretty good Inter-Jaccard indices (\lois{the indices are} distributed next to zero). 
Second, the DRAM latency PUF without filter has Intra-Jaccard indices distributed all over the spectrum (far from ideal), which does not satisfy the similarity property of a good quality PUF. \lois{This issue is solved by using the filtering mechanism}, \lois{which \minp{biases} \lois{the values of the Intra-Jaccard indices towards to one}, at the cost of increasing the evaluation time} (Section~\ref{sec:puf_evaluation_time}). 
Third, the DRAM Latency PUF has very good Inter-Jaccard indices, very close to zero. 
Fourth, the results from DDR3L \lois{chips} are better than those from DDR3 \lois{chips}, for both the DRAM Latency PUF and the UC-Dataplant PUF. We conclude that the \UPLAPUF{} is \lois{very effective on getting very similar of responses to the same challenge}, while maintaining uniqueness between responses from different memory segments.

Based on our results, a naive challenge-response authentication mechanism implemented with UC-Dataplant that correctly authenticates only when the response is exactly the expected \lois{(i.e., no filtering mechanism)}, has an average false rejection rate of 0.64\% and an average false acceptance rate of 0\%.

\begin{figure}[h] \centering
    \includegraphics[width=1.0\linewidth]{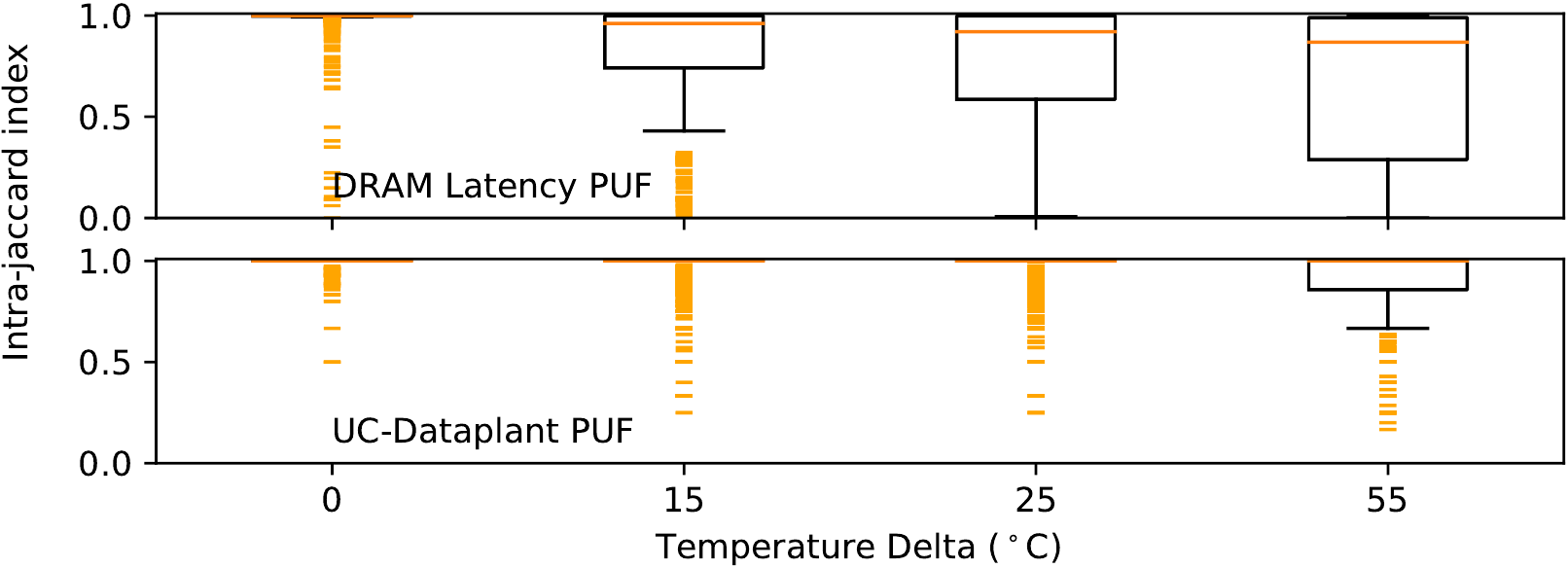}
    \caption{Intra-jaccard vs Temperature.}
    \label{fig:jaccard_temperature}
\end{figure}

\vspace{3pt}\noindent\textbf{Temperature and Aging Effects.}
\label{sec:temperature_effects}
To demonstrate how temperature affects the \lois{similarity of different responses to the same challenge}, we evaluate the \UPLAPUF{} and the DRAM latency PUF under different temperatures, ranging from 30$^{\circ}$C to 85$^{\circ}$C. \lois{We use the experimental setup from the previous experiment, a DRAM heater, and} a fine-grain temperature controller \lois{that can control the temperature with a precision of} $\pm$ 0.1$^{\circ}$C.
For this experiment, we only need to wait for 4 hours (instead of 48 hours), since cells discharge faster at high temperatures. 
Figure~\ref{fig:jaccard_temperature} shows the Intra-Jaccard indices between the \emph{same} segments under \emph{different} temperatures. Our main observation is that the \upla{} PUF is very robust to temperature changes, as \lois{the responses to the same challenge are very similar} even for extreme temperature changes (55$^{\circ}$C). The \lois{responses of the} DRAM latency PUF  \lois{are} much more \lois{sensitive to} temperature changes, confirming the results of the original work~\cite{kim2018}. 
\lois{We conclude that} the \upla{} PUF performs much better than the DRAM Latency PUF under changing temperature conditions.

\begin{figure}[h] \centering
    \includegraphics[width=0.7\linewidth]{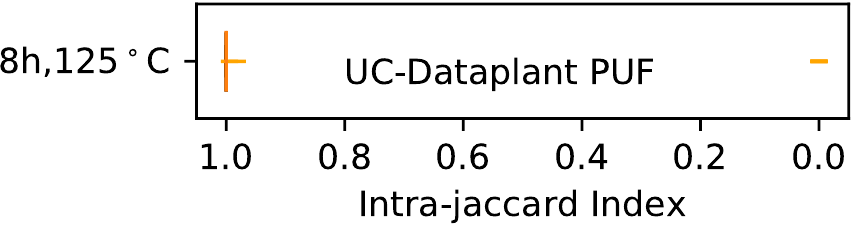}
    \caption{Intra-jaccard vs Accelerated aging during 8 hours at 125$\pmb{^\circ}$C.}
    \vspace{-4.5mm}
    \label{fig:jaccard_aging}
\end{figure}

To demonstrate how aging affects \lois{the similarity of different responses to the same challenge}, we use accelerated aging techniques to artificially age our DRAM \lois{chips}~\cite{Sabnis,Tehranipoor,Reynolds,Saha,Sonnenfeld}. We artificially age \lois{the} DRAM \lois{chips} by operating them at 125$^\circ$C degrees running stress tests during 8 hours. Figure~\ref{fig:jaccard_aging} shows the Intra-Jaccard indices between the same segments before and after the aging. We observe that UC-Dataplant is very robust to aging (most of the Jaccard indices are 1).

\subsubsection{Evaluation Time}
\label{sec:puf_evaluation_time}



\lois{The evaluation time of our \mechanism{} PUFs is in the same order of magnitude or faster than the fastest state-of-the-art DRAM PUF. The evaluation latency of the \UPLAPUF{} is the same latency as executing a precharge command, an activation command, and reading 8KB of data. The evaluation latency of \UESAPUF{}  is the same latency as performing an activation command, and reading 8KB of data.}


\lois{We observe that}, in the worst DRAM chip we tested, \mechanism{} PUF responses to the same challenge are exactly the same 99.72\% of the times. To ensure reliable PUF behavior, we also implement a filtering mechanism similar to the filter implemented in DRAM Latency PUF, but using only 5 \mechanism{} requests. Using this filtering mechanism, \emph{all} responses to the same challenge are exactly the same in our experiments. \lois{We also implement an alternative to the filtering mechanism (similar to PreLatPUF~\cite{talukder2019prelatpuf}) that profiles the DRAM memory and identifies the DRAM cells that provides robust PUF responses, which enables to eliminate the filtering mechanism.}
\lois{We compare our PUF to the same DRAM Latency PUF~\cite{kim2018} we use in the previous section, and to a PreLatPUF~\cite{talukder2019prelatpuf} that generates PUF responses by reducing the precharge latency.}
%
Table~\ref{table:latency_PUFs} summarizes the total evaluation time of the evaluated DRAM PUFs.




\begin{table}[h]
    \caption{PUF evaluation time of the DRAM Latency PUF, \lois{PreLatPUF,} and the \mechanism{} PUF, using 8KB memory segments.}
    \centering
    \footnotesize{}
    \setlength\tabcolsep{3pt} 
    \begin{tabular}{| p{1.4cm} | p{1.4cm} | p{1.4cm}| p{1.4cm} |}
      \hline
        {\bf Latency PUF} & {\bf \lois{PreLatPUF}} & {\bf \mechanism{} PUFs} &  {\bf \mechanism{} PUFs (no-filter)}  \\
        \hline
        \lois{88.2ms} & 1.59ms & 4.41 ms & 0.88 ms  \\
        \hline
    \end{tabular}
    \label{table:latency_PUFs}
\end{table}

We make two observations. First, the \mechanism{} PUFs with/without filter have 20x/100x lower evaluation latency than a DRAM Latency PUF. Second,  the \mechanism{} PUFs without filter are 1.8x faster than the PreLatPUF, but \mechanism{} PUFs with filter is 5x slower than PreLatPUF. \lois{Although the filter mechanism slows down the PUF evaluation latency, it also avoids other issues related to the no-filter mechanism, such as initial profiling, metadata accesses and management, etc.} 
We also perform experiments using a filtering mechanism with 5 responses in the DRAM Latency PUF, but this causes a large degradation of the PUF quality (Section~\ref{sec:emulating_upla}) and compromises \lois{the functionality and security of the PUF-based authentication protocol}.

\lois{We conclude that our mechanism is faster that the best state-of-the-art DRAM PUFs.}

\subsubsection{Randomness Analysis}
\label{sec:nist}

A secure key or seed should be random and have high-entropy. Although we already demonstrated the uniqueness of the responses between different memory segments (Section~\ref{sec:emulating_upla}), this does not guarantee properties \lois{such as} high-entropy. 

\vspace{3pt}\noindent\textbf{Methodology.} We analyze the randomness of the values generated by \upla{} with \emph{real} DRAM chips \lois{(Table~\ref{table:dram_chips})}, with the experimental setup of Section~\ref{sec:emulating_upla}. We generate a sequence of numbers composed by the relative position of the unpredictable values in a cache line. We use the NIST statistical test suite~\cite{rukhin2001statistical} to analyze the numbers generated by \upla{}.



\vspace{3pt}\noindent\textbf{Results.} We run the NIST test suite with the responses to different challenges from all the tested DRAM chips. We collect the PUF responses and we form up to 250KB sequence numbers. Table~\ref{table:NIST} shows the  average NIST p-values and NIST final results for the numbers generated by \upla{}. \lois{We use a customized version of the Von Neumann extractor~\cite{shaltiel2011introduction} for whitening the random stream.}




\begin{table}[h]
    \caption{\mechanism{} average results with the NIST randomness test suite. }
    \centering
    \footnotesize{}
    \setlength\tabcolsep{5pt} 
    \begin{tabular}{| c | c | c |}
      \hline
        {\bf NIST Test.} &  P-value & Result \\
        \hline
        monobit & 0.681 & PASS\\
        frequency\_within\_block & 1.000 & PASS\\
        runs & 0.298 & PASS\\
        longest\_run\_ones\_in\_a\_block & 0.287 & PASS\\
        binary\_matrix\_rank & 0.536 & PASS\\
        dft & 0.165 & PASS\\
        non\_overlapping\_template\_matching & 0.808 & PASS\\
        overlapping\_template\_matching & \lois{0.210} & \lois{PASS}\\
        maurers\_universal & 0.987 & PASS\\
        linear\_complexity & 0.0185 & PASS\\
        serial & 0.988 & PASS\\
        approximate\_entropy & 0.194 & PASS\\
        cumulative\_sums & 0.940 & PASS\\
        random\_excursion & 0.951 & PASS\\
        random\_excursion\_variant & 0.693 & PASS\\
        \hline
    \end{tabular}
    \label{table:NIST}
\end{table}

\lois{Our main observation is that} the numbers generated by \upla{} pass \lois{all} 15 NIST tests, which \lois{demonstrates} that our PUF is able to generate good quality random numbers. 




\subsection{Preventing Cold Boot Attacks}
\label{sec:eval_coldBootAttacks}

\lois{We} evaluate our new Command-Based Destruction and the Self-Destruction mechanisms described in Section~\ref{sec:coldbootattacks}. We customize the memory controller to implement the Command-Based Destruction with Rowclone, Lisa-clone, \uesa{}, and \upla{}. We implement Self-Destruction with \lois{\uesa{} and \upla{}}. We also implemented the TCG specification~\cite{computing2009tcg} for preventing cold boot attacks \lois{(see Section~\ref{sec:state-of-the-art_defenses})}.

\vspace{3pt}\noindent\textbf{Methodology.} We customize Ramulator~\cite{kim2016} to support the two proposed \mechanism{} implementations, Rowclone and Lisa-clone. Table~\ref{table:configuration} shows the summary of the DRAM and memory controller configurations used in our evaluation.

Our baseline is the TCG software cold boot attack prevention mechanism. We evaluate TCG by simulating a firmware approach that overwrites the memory with zeros by issuing regular write requests. To force writing back the data to memory from cache, we use an instruction that invalidates the data on cache (i.e., the \emph{CLFLUSH} instruction in x86). TCG does not require any hardware changes other than the BIOS customization.

We implement our Self-Destruction and Command-Based Destruction. Self-Destruction takes place entirely within DRAM, and it is implemented only with \mechanism{}. We implemented the two variants of Self-Destruction described in Section~\ref{sec:self-destruction}, namely Self-Destruction using self-refresh and Self-Destruction using burst refresh. Command-Based Destruction issues commands from the memory controller that destroy data in DRAM with  Rowclone, Lisa-clone, \lois{\uesa{}, and \upla{} primitives}.



To calculate latency of \mechanism{} we use the SA design described in~\cite{keeth2008dram}, and to calculate the energy we use a customized version of DRAMPower~\cite{chandrasekar2012drampower}. For \uesa{}, we use the same timing parameters as a regular activation, and for \upla{}, we use the same timing as a regular precharge (see Section~\ref{sec:dataplant} for details). 

{\setlength{\tabcolsep}{1pt}
\begin{table}[h]
    \footnotesize{}
    \caption{System configuration for evaluating our cold boot attack prevention mechanism.}
    \centering
    \fontsize{8}{11}\selectfont

    \begin{tabular}{ l l }
        \hline
        {\bf Proc.} & in-order core, 32KB L1 D\&I, 512KB L2 \\
        \hline
        {\bf Mem. Ctr.} &  64/64-entry read/write queue,  FR-FCFS~\cite{Rixner:2000,zuravleff1997controller}\\
        \hline
        {\bf  DRAM} &   1-2 channels, DDR3-1600 x8 11/11/11 \\
        \hline
    \end{tabular}
    \label{table:configuration}
\end{table}
}

We have already done a security analysis and hardware cost analysis of our mechanisms in Section~\ref{sec:discussion}. In this evaluation we show the latency improvements and the energy savings. 

\begin{figure}[h] \centering
    \includegraphics[width=1.0\linewidth]{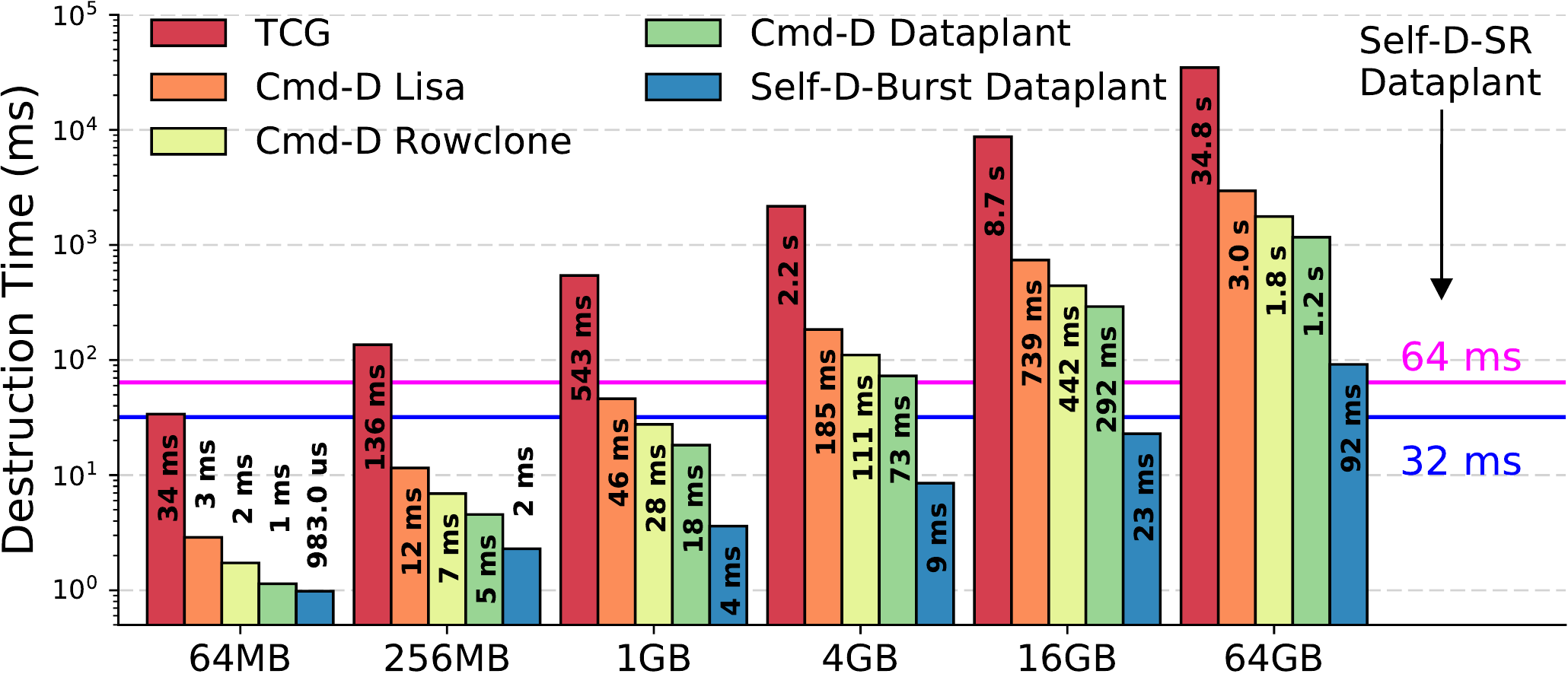}
    \caption{Time (log scale) to destroy all DRAM data using a software implementation (TCG), our Command-based Destruction (Cmd-D), our Self-Destruction using Burst refresh (Self-D-Burst Dataplant) and our Self-Destruction using Self-Refresh (Self-D-SR Dataplant).}

    \label{fig:time_destroy}
\end{figure}

\vspace{3pt}\noindent\textbf{\lois{Latency Results.}}
Figure~\ref{fig:time_destroy} shows the destruction time (in seconds, logarithmic scale) of the TCG software implementation, the Command-Based Destruction (Cmd-D) using all primitives, the Self-Destruction with Burst Refresh (Self-D-Burst) using \mechanism{}, and the Self-Destruction with Self-Refresh (Self-D-SR) using \mechanism{}.
\lois{We assume the same timing parameters for \uesa{} and \upla{} as regular activation and precharge commands (e.g., tFAW and tRDD) to meet internal DRAM power restrictions.} Although we show that \upla{} can perform faster than \uesa{} for individual primitives (Table~\ref{table:energy_latency}), the \lois{power} restrictions are very similar, which limits the throughput of \upla{}. In practice, the latency results of \uesa{} and \upla{} are identical for the cold boot attack prevention mechanism (Cmd-D Dataplant and Self-D-SR Dataplant in the figure).

We test different \lois{DRAM sizes}, from 64MB, used in memories specifically designed for IoT~\cite{HyperRAM}, to 64GB, used in high-end servers~\cite{64GB}. Our simulator takes into account all timing parameters defined by the DDR standard~\cite{jedec2012jedec}. The timing parameters for each size are taken from public datasheets released by vendors~\cite{micron}. For the memories that we don't have enough information about timing parameters (e.g., 64MB, 64GB), we extrapolate the parameters from \lois{existing} memory modules.


We make four major observations. 
\lois{First}, Self-Destruction based on burst refresh performs 19.5x better than Rowclone and 32.6x better than Lisa. 
Second, the Command-based destruction (Cmd-D) shows tolerable values for small and medium memory sizes. The Command-based Dataplant implementation performs 1.5x better than Rowclone, and 2.5x better than Lisa. 
\lois{Third}, Self-Destruction based on Self-Refresh has the same latency \lois{as} a regular refresh window. This \lois{approach} shows the best trade-off between performance and complexity (see Section~\ref{sec:self-destruction}). 
\lois{Fourth}, the software-based destruction mechanism (TCG) has a high latency, \lois{especially} for large DRAM sizes, which delays the boot time of the system significantly. \lois{Also, TCG does not provide strong security guaranties, as we discuss in Section~\ref{sec:state-of-the-art_defenses}}.


\vspace{3pt}\noindent\textbf{Energy Results.}
Table~\ref{table:energy_coldboot} shows the energy savings of \lois{our mechanism implemented with different hardware primitives,} compared to TCG.

\begin{table}[h]
    \caption{\lois{Energy savings of different mechanisms that destroy all DRAM data, compared to TCG.}}
    \centering
        \setlength\tabcolsep{1pt} 
    \footnotesize
    \begin{tabular}{| c | c | c | c |  }
      \hline
        {\bf Cmd-D Lisa} & {\bf Cmd-D Rowclone} & {\bf Self-D-Burst \& Self-D-SR Dataplant } \\

        \hline
        25x & 45x & 114x \\

        \hline
    \end{tabular}
    \label{table:energy_coldboot}
\end{table}

 The energy consumption of Self-Destruction is approximately the same as the Command-Based approach (excluding the energy of the bus). We observe that our \mechanism{} implementations show large energy savings compared to TCG (114x), and very significant energy savings compared to Lisa-clone (4.5x) and Rowclone (2.54x).




\vspace{3pt}\noindent\textbf{Comparison with other State-of-the-Art Mechanisms.}
There exist other mechanisms that protect against cold boot attacks \lois{that are fundamentally different to our approach}. This is the case with memory encryption, which provides strong security guarantees at the cost of additional energy consumption. Table~\ref{table:ColdBoot_comparison} shows the performance, power, and area overhead of our \lois{Self-Destruction mechanism} compared to ChaCha-8~\cite{Yitbarek2017} and AES-128~\cite{Yitbarek2017}, two low-cost ciphers that can be used to prevent cold boot attacks efficiently~\cite{Yitbarek2017}.

\begin{table}[h]
    \caption{Overhead of our Self-Destruction mechanism based on \mechanism{} compared to two other mechanisms to prevent cold boot attacks on an Intel Atom N280 processor.}
    \centering
    \begin{threeparttable}
    \footnotesize{}
    \setlength\tabcolsep{1pt} 
    \begin{tabular}{| c | c | c | c|}
      \hline
         & {\bf Self-Destruction}  & {\bf ChaCha-8 }~\cite{Yitbarek2017} & {\bf AES-128 }~\cite{Yitbarek2017}  \\
        \hline
        {\bf Runtime Performance} & 0\% & 0\% & 0\%\tnote{1}   \\
        \hline
        {\bf Runtime Power\tnote{2}} & 0\% & 17\% & 12\%   \\
        \hline
        {\bf Area} & $\sim$0\% & 0.9\% & 1.25\%   \\
        \hline
        
    \end{tabular}
    \begin{tablenotes}
     \item[1] when less than 16 back-to-back row hits.
     \item[2] at peak bandwidth utilization.
    \end{tablenotes}
    \end{threeparttable}
    \label{table:ColdBoot_comparison}
    \vspace{-2.5mm}
\end{table}

We make two main observations. First, our Self-Destruction mechanism has \emph{zero} performance and power overhead at runtime, and very low hardware cost, which make it difficult to beat as a low-cost method for preventing cold boot attacks. Second, although ChaCha-8 and AES-8 can be implemented for hiding the encryption latency in the common case~\cite{Yitbarek2017}, the power and area overheads of ChaCha-8 and AES-128 are significant in low-cost processors such as the Intel Atom N280. We conclude that our zero-overhead proposal is a very efficient way to protect against cold boot attacks in systems where encryption \footnote{While AES-128 and ChaCha-8 provide additional security features, we evaluate their ability to prevent cold boot attacks, as studied in recent literature~\cite{Yitbarek2017}.} is not an option.

\section{Related Work}
\label{sec:relatedWork}

To our knowledge, this is the first paper to propose \lois{low-cost} in-DRAM primitives that enable security mechanisms in \lois{all systems that use DRAM}.  We demonstrate two applications of our primitives: (1)~PUF-based authentication and (2)~cold boot attack prevention.


\vspace{5pt}\noindent\textbf{In-Memory Operations.}  We already compare \mechanism to Lisa~\cite{Chang2016LISA} and Rowclone~\cite{seshadri2013}. As we discussed, \mechanism{} is a very low overhead set of in-memory primitives to generate data for security mechanisms. Prior works on in-memory operations target other basic functionalities in commodity DRAM chips, such as AND/OR bitwise operations~\cite{seshadri2015, seshadri2017ambit, akerib2012using, akerib2017memory}. A number of works perform processing near memory using 3D-stacked memories, which often contain a logic layer, but such logic requires a much greater logic cost~\cite{Boroumand2017, Nai2017, Hsieh:2016:TOM, Hsieh:2016}.

\vspace{5pt}\noindent\textbf{PUFs.}
\lois{We have already compared our \mechanism{} PUFs to the DRAM Latency PUF~\cite{kim2018} \lois{and to the PreLatPUF~\cite{talukder2019prelatpuf}}.}
Many PUFs have been investigated in different components, such as SRAM~\cite{holcomb2007initial,holcomb2009power, Bhargava2012, zheng2013, xiao2014, Bacha2015}, ASIC logic~\cite{Daihyun2005, vanderLeest:2010}, and DRAM~\cite{Tehranipoor:2015, Sutar2016, Xiong2016, kim2018, talukder2019prelatpuf}.
\lois{There is one DRAM PUF that can be accessed during runtime (other than the DRAM Latency PUF)}. The Runtime DRAM PUF~\cite{Xiong2016} disables refresh in certain memory regions that are initialized with specific values. The PUF response is a function of the errors produced in the cells due to a lack of refresh after some time \emph{t}. 
Our \mechanism-based PUFs have lower evaluation times than the state-of-the-art DRAM PUFs. \lois{In addition, our \mechanism{} PUFs provide 1) non-destructive read-out, 2) low evaluation latency, 3) robust responses, \lois{4) resiliency to temperature changes, and 5)  data-independent responses, characteristics} that any other DRAM PUFs can provide \lois{all together}.}

\vspace{5pt}\noindent\textbf{Cold Boot Attacks}
Several works propose encryption mechanisms to protect data against different attacks, including cold boot attacks~\cite{arnold2004ibm, Yang:2005, Duc2006, Henson:2014}, which usually introduce performance and energy overheads. Various proposals attempt to reduce these overheads~\cite{Suh:2003, Yang:2005, Yan2006, Rogers2007}. Intel's Software Guard Extensions (SGX)~\cite{costan2016intel} can create protected areas of memory that ensure confidentiality and integrity of the data by using strong encryption (AES) and message authentication codes (MAC). Other papers propose to use modern stream ciphers as a fast way to encrypt memory~\cite{Yitbarek2017, bernstein2008chacha}.

A work on data lifetime management~\cite{lee2017dram} proposes to disable access to the data in DRAM, as another solution for cold boot attacks. The authors provide a new flag in the DRAM decoder, controlled by a DRAM command, that controls the access to a DRAM row for untrusted programs. Unlike our \mechanism-based cold boot attack prevention mechanism, this prior work does not prevent an attacker with \emph{physical access} from having free access to the rows.

Seol et al.~\cite{Seol2017} propose a mechanism to initialize DRAM with a reset operation based on connect/disconnect power lines. This reset operation has larger \lois{latency} than \mechanism{}, \lois{as it requires} a precharge and an activation command. In comparison, \mechanism{} requires \lois{only one command to destroy the content of the cell}. Memory scramblers are the main protection against cold boot attacks in modern \emph{unencrypted} memory systems. However, these scrambling mechanisms are not sufficient to protect against cold boot attacks, \lois{as demonstrated by prior works}~\cite{BAUER2016S65, Yitbarek2017}.

Our mechanism for protecting cold boot attacks improves the state-of-the-art by proposing a very simple mechanism with no performance or energy overhead at runtime.

%



\section{Conclusion}

We propose \emph{\mechanism}, a set of low-cost, highly efficient, and reliable \lois{in-DRAM} primitives that can enable important security mechanisms \lois{at low-cost} on any device that \lois{uses} DRAM. The main idea of \mechanism{} is to slightly modify the internal DRAM timing signals to expose the inherent process variation found in all DRAM chips for generating \emph{unpredictable} but reproducible values within DRAM. 
We \lois{build} two low-cost security mechanisms using \mechanism{} \lois{for demonstrating the potential of our primitive}. \lois{First, a \mechanism{}-based physically unclonable functions (PUFs) that improve state-of-the-art DRAM PUFs.
Second, a new cold boot attack prevention mechanism based on \mechanism{} that has zero performance and energy overhead at runtime.}

\lois{We show in our evaluation} that these two mechanisms are significantly faster and more energy-efficient than their state-of-the-art counterparts, with the same \lois{or better} security guarantees.
We conclude that \mechanism{} can effectively enable low-cost and low-power security mechanisms for all types of devices that \lois{use} DRAM, \lois{from low-cost devices} to high-end servers. This paper is a first step towards a more secure DRAM memory. We hope and expect that the availability of \mechanism{} in commodity DRAM chips will enable new security features and applications. 

\bibliographystyle{ieeetr}
\bibliography{main}

\begin{appendices}
\section{\dtran{} primitive}
\label{sec:deterministic_dataplant}

\lois{In this Appendix we discuss \dtran{} (\underline{D}eterministic \mechanism{}), an alternative \mechanism{} implementation with different characteristics than \uesa{} and \upla{}}. \dtran{} generates \lois{\emph{deterministic}} values within the SAs by adding a transistor to each SA in the row buffer. The generated values can be stored in DRAM, or can be read by the processor from the SAs without overwriting data in DRAM.

\begin{figure}[h] \centering
    \includegraphics[width=0.83\linewidth]{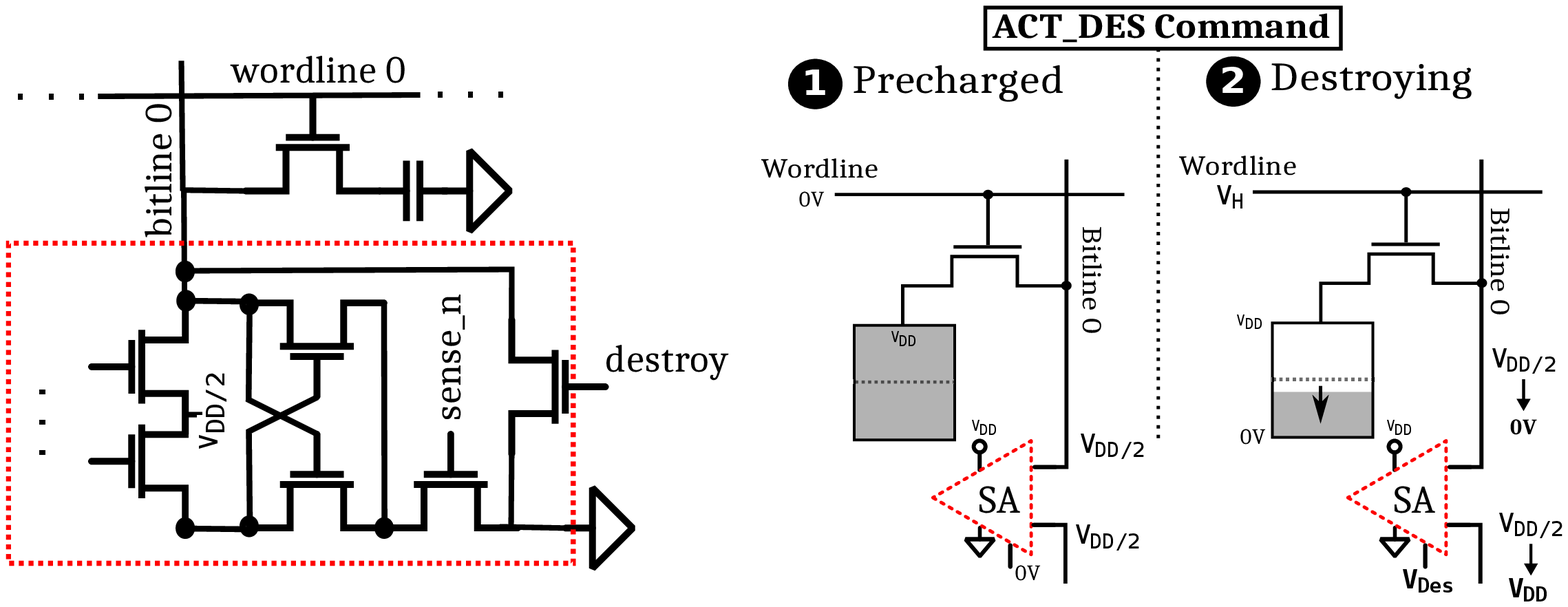}%
    \caption{\mechanism{} transistor placement (left), and behaviour (right).}
    \vspace{-1mm}
    \label{fig:activation_destroy}
\end{figure}

\dtran{} \emph{deterministically} drives the bitline voltage level to zero (0V) or one ($V_{dd}$), and optionally writes the generated value into the cell. The key idea is to add an additional path connecting a fixed voltage level to the bitline. To this end, we add an additional transistor controlled by a \emph{Dplant} signal. Figure~\ref{fig:activation_destroy} (left) shows how this transistor is connected into the SA to generate a "zero" or a "one" value. Figure~\ref{fig:activation_destroy} (right) illustrates how the Dataplant transistor drives the cell towards a deterministic value (zero in this case).


\begin{figure}[h]
    \centering
    \includegraphics[width=0.65\linewidth]{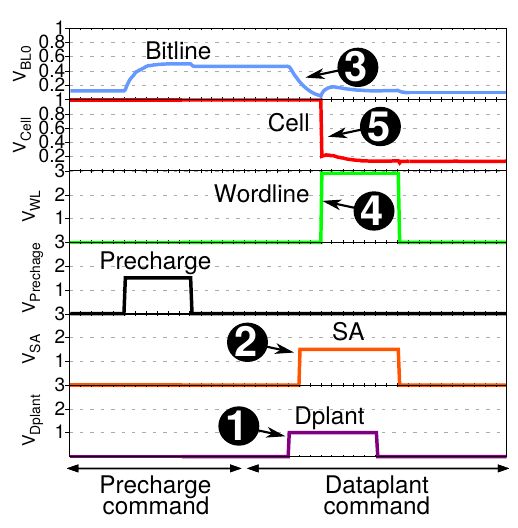}
    \vspace{-1.5mm}
    \caption{SPICE simulation of the internal DRAM signals involved in \dtran{}. $V_{dd}$=1V and the original content of the DRAM cell is ``one" ($V_{Cell}$).}
    \vspace{-1.5mm}
    \label{fig:dtran_act_timing2}
\end{figure}

Figure~\ref{fig:dtran_act_timing2} shows how the value is generated (including the optional overwriting of the cell). First, the \emph{Dplant} ($V_{Dplant}$) \circled{1}  and the SA \circled{2}  ($V_{SA}$) signals are triggered for driving the bitline to the deterministic voltage level \lois{(zero in the example)} \circled{3}. Then, if the wordline is triggered \circled{4}  ($V_{WL}$), the generated value is moved to the DRAM cell \circled{5}  ($V_{Cell}$), overwriting the previous content of the cell. 

\subsection{Evaluation}
Table~\ref{table:energy_latency_dplant} shows the absolute \lois{value} and the reduction of latency and energy of \lois{\dtran{}}  when generating a value of 8\,KB, \lois{compared to the baseline, Rowclone, Lisa, \uesa{} and \upla{}}. \lois{We also show the in-DRAM latency and energy consumption of standard activation and precharge commands.} The table also shows the \mechanism{} energy breakdown (value generation + overwriting) of \dtran{}  (\dtranShort{}). The energy consumption of \dtran{} is very similar to \uesa{} and \upla{} because the additional transistor of \dtran{} has a very low effect on the energy consumption. 

\begin{table}[h]
    \caption{Latency and energy of different primitives for \lois{generating 8KB of data}.}
    \centering
    \footnotesize
    \begin{tabular}{| r | c | c | c | c | c |}
      \multicolumn{1}{c}{} & \multicolumn{2}{c}{ {\bf Absolute}} & \multicolumn{2}{c}{{\bf Reduction}}\\
      \hline
         {\bf Primitive } & {\bf Lat. (ns)} & {\bf Ener. (nJ)} & {\bf Lat.} & {\bf Ener.}\\
        \hline
        \hline
        Baseline & 546 & 2000 & 1.0x & 1.0x \\
        \hline
        Lisa-clone & 148.5 & 90 & 3.67x & 22.2x \\
        \hline
        Rowclone & 90 & 50 & 6.06x & 41.5x \\
        \hline
        \hline
        \lois{Activation} & 35 & 17.3 & 15.6x & 116x\\
        \hline
        \lois{Precharge} & 13 & 17.2 & 42x & 116x\\
        \hline
        \hline
        \uesaShort{} & 35 & 17.3 = 7.3 + 10 & 15.6x & 116x \\
        \hline
        \uplaShort{} & 13 & 17.2 & 42x & 116x\\
        \hline
        {\bf \dtranShort{}} & 35 & 18 = 8 + 10 & 15.6x & 111x \\
        \hline
    \end{tabular}
    \label{table:energy_latency_dplant}
\end{table}

\vspace{3pt}\noindent\textbf{Area Overhead.} 
The \lois{main area} overhead of \dtran{} is \lois{caused by} an additional transistor per SA. Our SPICE simulations show that this transistor can be very small, but we consider a full-size transistor to avoid possible layout and fabrication issues. Considering an SA composed of 20 transistors~\cite{keeth2008dram}, the worst case overall area overhead is between 0.4\% and 2\% depending on the DRAM design. \footnote{This depends on the number of cells per bitline in the subarray, which determines the total number of SAs in the module.}.
\section{Secure Deallocation}
\label{sec:secureDeallocationEval}
\label{sec:eval_dealloc}


\label{sec:othermechanisms}

\lois{In this Appendix we describe and evaluate secure deallocation, a security application that can be efficiently implemented with any \mechanism{} primitive. However, this application is especially suitable for \dtran{} (Appendix~\ref{sec:deterministic_dataplant}) because most Operating Systems require that newly allocated memory is filled with zero values.}


Today's applications, especially web servers, web browsers, and word processors, do not immediately remove data from memory when it is no longer needed. Instead, the data is physically erased only when the memory is required for other uses. As a consequence, sensitive data could remain in memory for an indefinite amount of time, which augments the risk of exposure.

Secure deallocation~\cite{chow2005shredding, Anikeev:2013, Sha2018, harrison2007protecting, geambasu2009vanish, Garfinkel:2004} is a technique that set the data to zero at the moment of deallocation, or as soon as the data is not needed anymore. This technique reduces the time that critical data is exposed to attacks. Vanish~\cite{geambasu2009vanish} proposes a similar idea in which the old copies of data are self-destroyed after a specific amount of time. \mechanism{} enables the implementation of the previous techniques with very low latency, energy, and area overhead. 

\subsection{Evaluation}

\vspace{3pt}\noindent\textbf{Methodology.} We simulate \uesa{} \lois{(Section~\ref{sec:u-esa})}, \upla{} \lois{(Section~\ref{sec:u-pla})}, \dtran{} (Appendix~\ref{sec:deterministic_dataplant}), Lisa-clone~\cite{Chang2016LISA}, Rowclone~\cite{seshadri2013} and a software secure deallocation mechanism ~\cite{chow2005shredding}. 

We \lois{customize} Ramulator~\cite{kim2016} to support all the mechanisms on in-order cores. To generate the traces that drive our simulator, we use PIN~\cite{Luk:2005} for user-level traces, and \lois{the Bochs~\cite{bochs}} full-system emulator to generate the memory traces that include Linux kernel page allocations and deallocations. 

To calculate the area, energy, and latency of our \mechanism{} \lois{primitives}, we use the SA design described in Section~\ref{sec:dataplant}. 
To estimate the energy \lois{consumption of the DRAM module}, we use a customized version of DRAMPower~\cite{chandrasekar2012drampower}. Table~\ref{table:configuration_global} shows the system configuration used \lois{in} our evaluation.

{\setlength{\tabcolsep}{1pt}
\begin{table}[h]
    \centering
    \scriptsize{}
    \caption{System configuration.}\label{table:configuration_global}
    \fontsize{8}{11}\selectfont
    \begin{tabular}{ l l }
        \hline
        {\bf Processor} &  1-4 cores, in-order, \\ [0.5ex]
        \hline
        {\bf Cache} &  L1:64KB, L2:512KB per core, 64B lines\\
        \hline
        {\bf Mem. Ctr.} &  64/64-entry read/write queue,  FR-FCFS~\cite{Rixner:2000,zuravleff1997controller}\\
        \hline
        {\bf  DRAM} &   1 channel, DDR3-1600 x8 11/11/11 \\
        \hline
    \end{tabular}
    
\end{table}
}

\lois{Table~\ref{table:benchmarks_init} describes the \lois{6} memory-allocation-intensive benchmarks we use. For the multicore evaluation (4 cores), we choose 50 mixes of workloads, in which each mix is composed by two memory-allocation-intensive benchmarks, and two benchmarks that are non-memory-allocation-intensive. The non-memory-allocation-intensive benchmarks are TPC-C~\cite{tpc}, TCP-H~\cite{tpc}, STREAM~\cite{stream}, SPEC2006~\cite{spec2006}, DynoGraph (pagerank, bfs, stream)~\cite{dynograph}, and HPCC RandomAccess~\cite{hpc}. Table~\ref{table:multicore_benchmarks} shows 5 representative benchmark mixes.}

\begin{table}[h]
    \centering
    \caption{\lois{Memory-allocation-intensive benchmarks used} for evaluating secure deallocation.}
    \scriptsize{}
    \begin{tabular}{| r | l |}
      \hline
        {\bf Bench.} & {\bf Description} \\
        \hline
        \hline
        mysql &  MySQL~\cite{mysql} loading the sample {\it employeedb}.\\
        \hline
        mcached &  Memcached~\cite{memcached}, a memory object caching system \\
        \hline
        compiler &  Compilation phase from the GNU C compiler \\
        \hline
        bootup &  Linux kernel booting up phase \\
        \hline
        shell &  Script running 'find' in a directory tree with 'ls' \\
        \hline
        malloc & stress-ng~\cite{stress-ng} stressing the malloc primitive \\
     \hline
    \end{tabular}
    \label{table:benchmarks_init}
\end{table}

\begin{table}[h]
    \centering
    \caption{Five representative mixes (out of 50) used in the multicore evaluation for secure deallocation.}
    \scriptsize{}
    \setlength\tabcolsep{1.5pt}
    \begin{tabular}{ r  l r l}
      \hline
        {\bf MIX1:} & malloc, bootup, tpcc64, libquantum & {\bf MIX4:} & malloc, shell, xalancbmk, bzip2 \\
        \hline
        {\bf MIX2:} & shell, bootup, lbm, xalancbmk & {\bf MIX5:} &malloc, malloc, astar, condmat\\ %
        \hline
        {\bf MIX3:} & bootup, shell, pagerank, pagerank &&\\
        \hline
    \end{tabular}
    \label{table:multicore_benchmarks}
\end{table}

\begin{figure}[h] \centering
    \includegraphics[width=1.0\linewidth]{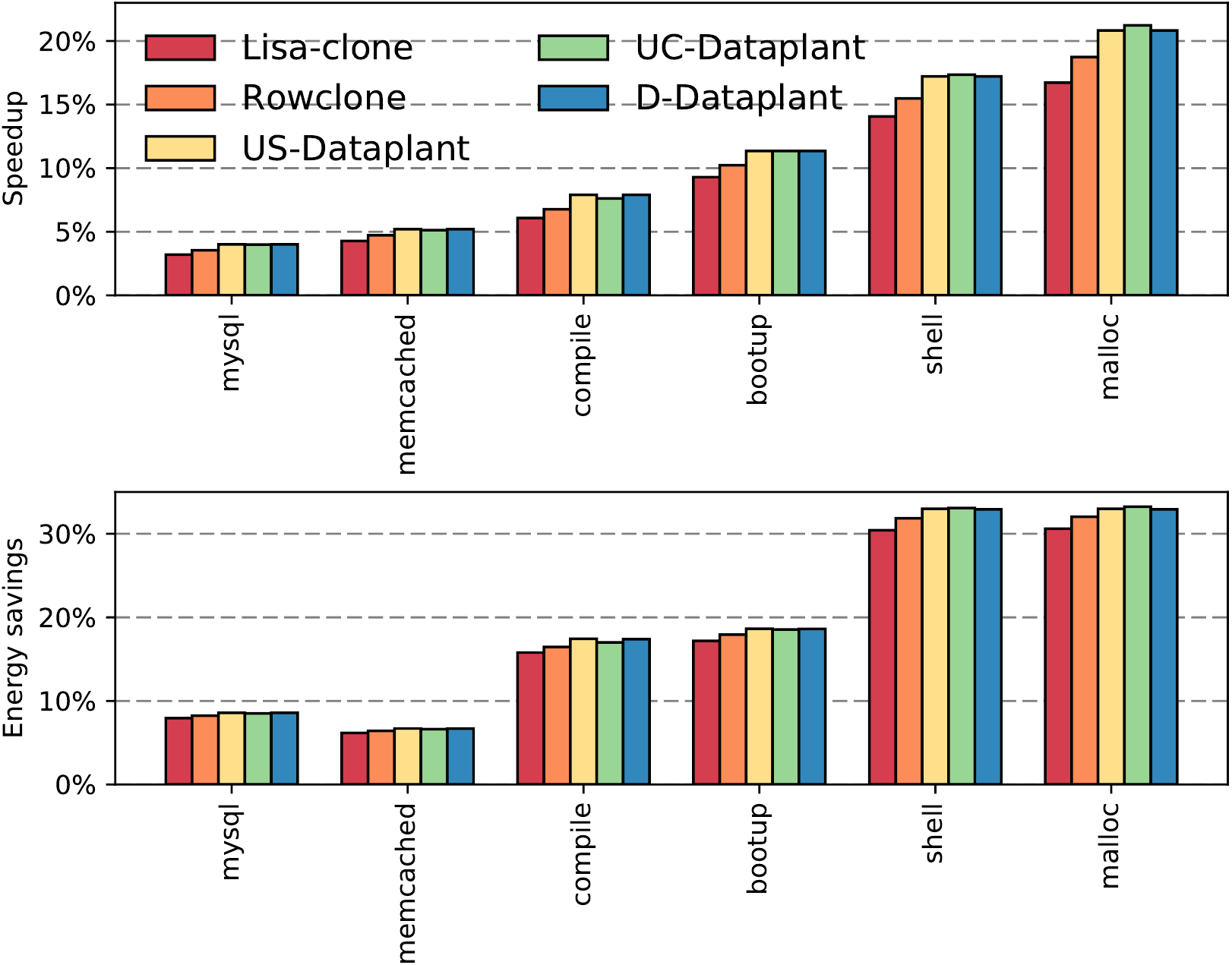}
    \caption{Single core speedup (larger is better) and energy savings (larger is better) of the secure deallocation hardware approaches compared to a software approach.}
    \label{fig:single_core}
\end{figure}

\vspace{3pt}\noindent\textbf{Results.}
Figure~\ref{fig:single_core} shows the \lois{single-core speedup} (higher is better) and energy savings (\lois{higher} is better) of \mechanism{} and other state-of-the-art mechanisms (Rowclone and Lisa-clone) normalized to a software secure deallocation implementation. 

\lois{We make three observations}. First, \lois{all} hardware implementations improve the performance up to  21\% and the energy savings up to 34\%, compared to a software implementation. Second, \mechanism{} performs better than Lisa-clone and Rowclone in all cases, both in performance and energy consumption. Third, the performance improvements and energy savings of \mechanism{} compared to Rowclone and Lisa-clone are not very large for some benchmarks. Note, however, that our approaches are much easier to integrate on commodity DRAM chips than Lisa-clone or Rowclone (Section~\ref{sec:dataplant}). 

\begin{figure}[h] \centering
    \includegraphics[width=1.0\linewidth]{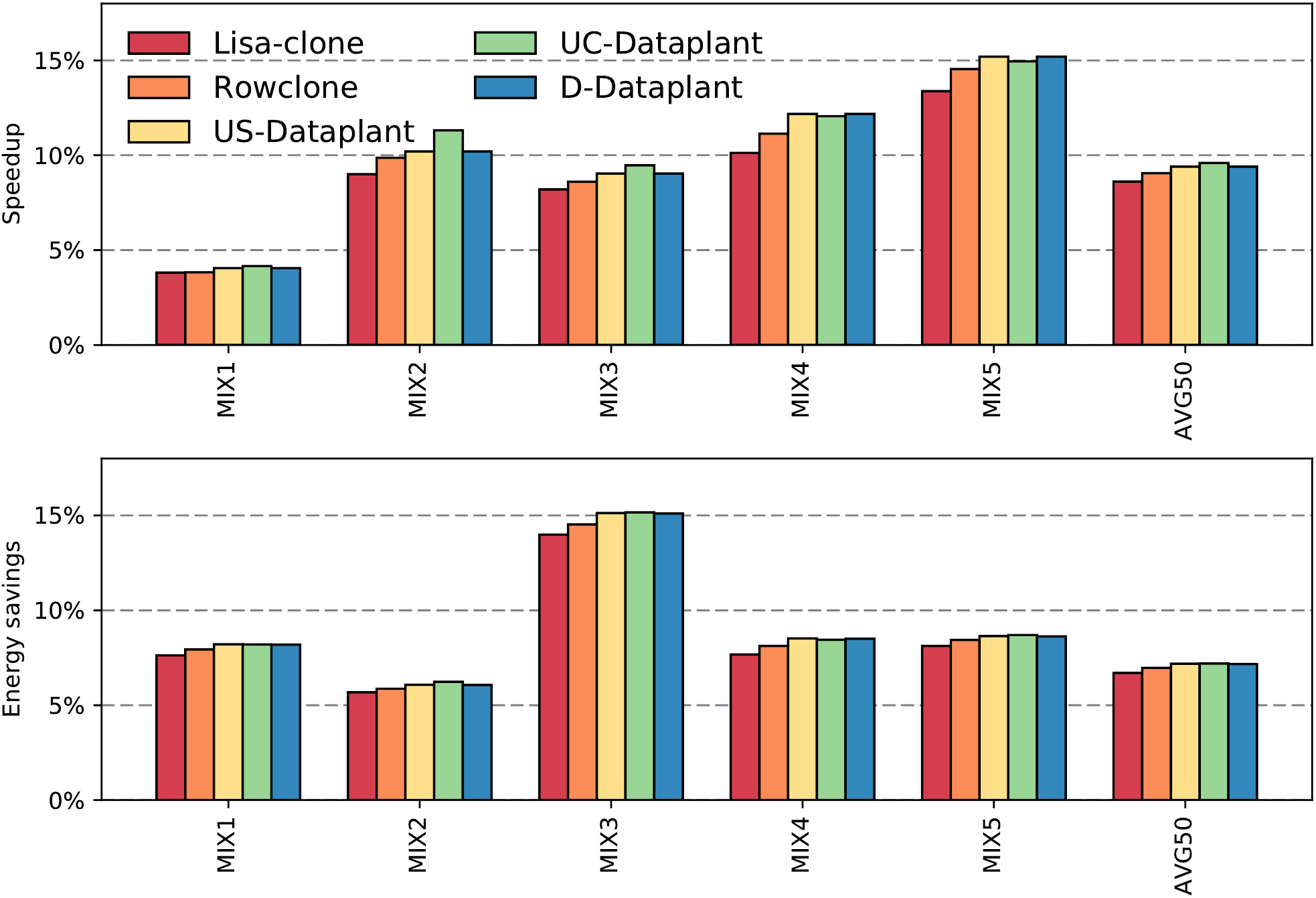}
    \caption{4 core speedup (larger is better) and energy savings (larger is better) of the secure deallocation hardware approaches compared to a software approach.}
    \label{fig:multi_core}
\end{figure}

\lois{Figure~\ref{fig:multi_core} shows the \lois{speedup} and energy savings of \mechanism{} and other state-of-the-art mechanisms in a 4-core processor, normalized to a software secure deallocation implementation}. 

\lois{We make the same observations \lois{as} for a single core processor: 1) all hardware approaches improve the software implementation, 2) \mechanism{}  performs better than Lisa-clone and Rowclone, and 3) the performance improvements and energy savings of \mechanism{} compared to Rowclone and Lisa-clone are not very large.}


\lois{We conclude that 1) implementing secure deallocation with hardware approaches can have significant performance and energy benefits, and 2) \mechanism{} is the best option to accelerate secure deallocation because it has the best performance and energy results, and it is the approach that is simpler to integrate in commodity DRAM chips.}

 \section{DRAM Logic}
\label{sec:DRAM_logic}
 
 To the best of our knowledge, there is no public information about how vendors implement the control logic, or what is the specific circuit design of that logic. However, this logic should be very similar to other DRAM logic. There are at least two types of logic styles that can be used to issue signals in DRAM~\cite{keeth2008dram}. First, the delay-chain logic style is composed of a mixture of combinational gates, monostable multivibrators, and latches. This logic is implemented with a series of delays such that a sequence of functions can be executed without a clock. Second, the domino logic uses logic gates that require a clock signal, and it enables to implement logic functions from a cascade of events. The hardware cost of our primitives is very low in both cases, because our mechanism can reuse most of the logic for generating the activation and precharge timing signals.
\end{appendices}

\end{document}